\documentstyle[emulateapj,apjfonts]{article}

\def\mkfigbox#1#2{
\hbox{ \epsfxsize=#2 \epsfbox{#1} \relax} }

\def\Fig{{\footnotesize {\sc Fig.\/}~\,\thefigure. (continued) ---}}

\def\kms{km s$^{-1}$} 
\def\etal{{\it et al.}}

\def\fpparam#1{\hbox{$r_e\langle I\rangle_e^{#1}$}}

\def\Sec{${}^{\prime\prime}$\llap{.}}

\def\etal{{\it et~al.\/}}

\def\kms{{km~s$^{-1}$}}
\def\kpc-1{{kpc$^{-1}$}}
\def\Mpc-1{{Mpc$^{-1}$}}
\def\s-1{{sec$^{-1}$}}
\def\pdeg2{{deg$^{-2}$}}

\def\h0{{H$_0$}}
\def\q0{{$q_0$}}

\def\rms{{\it rms\/}}

\def\etal{{\it et al.\/}}
\def\kms{\hbox{$\rm km\,s^{-1}$}}

\def\ltsima{$\scriptscriptstyle \; \buildrel < \over \sim \;$}
\def\simlt{\lower.3ex\hbox{\ltsima}}

\def\gtsima{$\scriptscriptstyle \; \buildrel > \over \sim \;$}
\def\simgt{\lower.3ex\hbox{\gtsima}}

\def\about{\raise.3ex\hbox{$\scriptscriptstyle \sim $}}

\def\Sec{\hbox{${}^{\prime\prime}$\llap{.}}}

\def\sqr#1#2{{\vcenter{\vbox{\hrule height.#2pt
        \hbox{\vrule width.#2pt height#1pt \kern#1pt
        \vrule width.#2pt}
        \hrule height.#2pt}}}}

\def\dns{\hbox{$D_n$-$\sigma$}}

\lefthead{Kelson \etal}
\righthead{Key Project}
\slugcomment{Accepted for publication in the Astrophysical Journal}

\begin{document}

\title{THE EXTRAGALACTIC DISTANCE SCALE KEY PROJECT XXVII. A
DERIVATION OF THE HUBBLE CONSTANT USING THE FUNDAMENTAL PLANE AND
\dns\ RELATIONS IN LEO I, VIRGO, AND FORNAX}

\author{Daniel D. Kelson$^1$, Garth D. Illingworth$^2$, John L.
Tonry$^3$, Wendy L. Freedman$^4$, Robert C. Kennicutt, Jr.$^5$, Jeremy
R. Mould$^6$, John A. Graham$^1$, John P. Huchra$^7$, Lucas M.
Macri$^7$, Barry F. Madore$^8$, Laura Ferrarese$^{9}$, Brad K.
Gibson$^{10}$, Shoko Sakai$^{11}$, Peter B. Stetson$^{12}$, Edward A.
Ajhar$^{11}$, John P. Blakeslee$^{9}$, Alan Dressler$^4$, Holland C.
Ford$^{13}$, Shaun M.G. Hughes$^{14}$, Kim M. Sebo$^6$, and Nancy A.
Silbermann$^8$}

\altaffiltext{1}{Department of Terrestrial Magnetism, Carnegie Institution
of Washington, 5241 Broad Branch Rd., NW, Washington, DC 20015}

\altaffiltext{2}{University of California Observatories / Lick Observatory,
Board of Studies in Astronomy and Astrophysics, University of California,
Santa Cruz, CA 95064}

\altaffiltext{3}{Inst. for Astronomy, 2680 Woodlawn Dr., Honolulu, HI
96822}

\altaffiltext{4}{Carnegie Observatories, Pasadena CA 91101}

\altaffiltext{5}{Steward Observatories, University of Arizona, Tucson AZ
85721}

\altaffiltext{6}{Research School of Astronomy \& Astrophysics,
Institute of Advanced Studies, Australian National University, ACT
2611, Australia}

\altaffiltext{7}{Harvard-Smithsonian Center for Astrophysics, Cambridge,
MA 02138}

\altaffiltext{8}{IPAC, California Institute of Technology, Pasadena CA
91125, USA}

\altaffiltext{9}{Department of Astronomy, California Institute of
Technology, Mail Stop 105-24, Pasadena CA 91125}

\altaffiltext{10}{Center for Astrophysics \& Space Astronomy, University of
Colorado, Boulder CO 80309-0389}

\altaffiltext{11}{National Optical Astronomical Observatory, Tucson, AZ 85726}

\altaffiltext{12}{Dominion Astrophysical Observatory, Victoria, BC V8X 4M6,
Canada}

\altaffiltext{13}{John Hopkins University and Space Telescope Institute,
Baltimore MD 21218}

\altaffiltext{14}{Royal Greenwich Observatory, Cambridge CB3 0EZ, UK;
Current address: Institute of Astronomy, Madingley Road, Cambridge,
CB3 0HA, UK}

\begin{abstract}

Using published photometry and spectroscopy, we construct the
fundamental plane and $D_n\hbox{-}\sigma$ relations in Leo I, Virgo and
Fornax. The published Cepheid P-L relations to spirals in these clusters
fixes the relation between angular size and metric distance for both the
fundamental plane and $D_n\hbox{-}\sigma$ relations. Using the locally
calibrated fundamental plane, we infer distances to a sample of clusters
with a mean redshift of $cz \approx 6000$ \kms, and derive a value of
$H_0=78\pm 5\pm 9$ \kms\Mpc-1 (random, systematic) for the local
expansion rate. This value includes a correction for depth effects in
the Cepheid distances to the nearby clusters, which decreased the
deduced value of the expansion rate by $5\% \pm 5\%$. If one further
adopts the metallicity correction to the Cepheid PL relation, as derived
by the Key Project, the value of the Hubble constant would decrease by a
further $6\%\pm 4\%$. These two sources of systematic error, when
combined with a $\pm 6\%$ error due to the uncertainty in the distance
to the Large Magellanic Cloud, a $\pm 4\%$ error due to uncertainties in
the WFPC2 calibration, and several small sources of uncertainty in the
fundamental plane analysis, combine to yield a total systematic
uncertainty of $\pm 11\%$. We find that the values obtained using either
the CMB, or a flow-field model, for the reference frame of the distant
clusters, agree to within 1\%. The \dns\ relation also produces similar
results, as expected from the correlated nature of the two scaling
relations. A complete discussion of the sources of random and systematic
error in this determination of the Hubble constant is also given, in
order to facilitate comparison with the other secondary indicators being
used by the Key Project.

\end{abstract}



\section{Introduction}

The goal of the distance scale Key Project is to measure distances to
galaxies where the Hubble flow is expected to dominate over local
velocity perturbations, and derive a value of the Hubble constant
accurate to $\pm 10\%$. However, much of the controversy over the
Hubble constant ($H_0$) arises from disagreements over the secondary
distance indicators which are used (see, for example, Jacoby {\it et
al.\/} 1992, Fukugita, Hogan, \& Peebles 1993). Therefore, a concerted
effort is required to define zero points for a variety of independent
indicators, including, for example, the Tully-Fisher relation (Tully
\& Fisher 1977, Aaronson, Mould, \& Huchra 1979), \dns\ (Dressler
\etal\ 1987, Lynden-Bell \etal\ 1987), surface brightness fluctuations
(Tonry \& Schneider 1988), and methods based on supernovae (cf. Riess,
Press, \& Kirshner 1996). By calibrating several distance indicators,
understanding their systematic differences, and combining them in a
sensible way, it is hoped that the stated goal of 10\% accuracy may be
achieved (Mould \etal\ 1999a).

In this paper, we use data from the literature to construct the
fundamental plane (FP) and \dns\ relations in Leo I, Virgo, and
Fornax. A number of Cepheid distances have been published by the Key
Project to Leo I, Virgo, and Fornax. These groups, or clusters have
several early-type galaxies for which kinematic and photometric data
exist in the literature. These data, with the Cepheid distances, allow
us to calibrate the FP and \dns\ relations, and thus use these
relations as secondary distance indicators in the distance scale
ladder. Using the fundamental plane and \dns\, one can derive the
distance to Coma and several other distant clusters, for which the
peculiar velocities are expected to be a small component of the
observed radial velocities, and accurately determine the local
expansion rate.

The results of this work can be directly compared to Sakai \etal\
(1999) who calibrate the local Tully-Fisher relation (Tully \& Fisher
1977), and use it to measure distances to clusters in the flow-field;
Ferrarese \etal\ (1999), who calibrate other Population II indicators
such as the planetary nebula luminosity function, and the surface
brightness fluctuation method; and Gibson \etal\ (1999), who
recalibrate the Type Ia supernovae method. The values of $H_0$ from
each of these secondary indicators are combined with the results from
the fundamental plane and \dns\ by Mould \etal\ (1999a) and Freedman
\etal\ (1999).

This paper is structured as follows. In \S \ref{fpintro}, we give a
brief introduction to the fundamental plane and \dns\ scaling
relations. Section \ref{data} outlines the data taken from the
literature, as well as the subsequent treatment required to ready the
data for fundamental plane analysis. We discuss the Cepheid distances
to Leo I, Virgo, and Fornax in \S \ref{adopt}. In \S \ref{fplocal} the
fundamental plane relations in Leo I, Virgo, and Fornax are compared
to that in Coma and in other distant clusters and a value for the
Hubble constant is derived. Section \ref{dnsigma} is devoted to the
analysis of the local \dns\ relation, with its implied value of the
Hubble constant, and a comparison of the \dns\ relation with the FP.
Following a brief discussion of the scatter in the local fundamental
plane and \dns\ relations in \S \ref{scatter}, we discuss systematic
corrections to the derived value of $H_0$ (\S \ref{syscor}).  A final
value for the Hubble constant is given in \S \ref{errors}, along with
a detailed error budget and an estimate of the total uncertainty.
Lastly, a brief comparison of our data with previous work in the
literature is presented in Section \ref{complit}.


\section{The Fundamental Plane and \dns\ Relations as Distance
Indicators}
\label{fpintro}

Elliptical galaxies appear to be a family of objects described by a
small number of physical parameters. Because of dynamical equilibrium
and the virial theorem, one expects that the characteristic internal
velocities of elliptical galaxies should be correlated with their
binding energies, and assuming light traces mass in a similar way for
all elliptical galaxies, one therefore predicts a correlation between
$\sigma$, and luminosity. In this way, the Faber-Jackson relation of
elliptical galaxies (Faber \& Jackson 1976) is analogous to the
Tully-Fisher relation for spirals: The Faber-Jackson relation allowed
for early-type galaxies to be used as standard candles for measuring
distances to massive clusters, measuring the Hubble constant, and
mapping the peculiar velocity field.

The introduction of surface brightness by Dressler \etal\ (1987) was a
major improvement to the Faber-Jackson relation, reducing the scatter
in elliptical galaxy scaling relations by 50\%. The \dns\ relation
blended the observables of galaxy size and surface brightness by
defining $D_n$, the diameter within which the surface brightness was
equal to some fixed value. For galaxies with similarly shaped growth
curves, $D_n$ is a well-defined function of half-light diameter,
$D_e$, and mean surface brightness, $\langle I\rangle_e$, within
$D_e$, such that $D_n = f(D_e,\langle I\rangle_e)$.

Simultaneously, Djorgovski \& Davis (1987) reported a ``fundamental''
relation between $r_e$, $\sigma$, $\langle I \rangle_e$. Both groups
speculated that \dns\ was simply an edge-on projection of the new
``fundamental plane:''
\begin{equation}
r_e \propto \sigma^\alpha \langle I\rangle_e^\beta
\end{equation}
where $\alpha\approx 1.2$ and $\beta\approx -0.85$ (Djorgovski \&
Davis 1987). Lucey, Bower, \& Ellis (1991) later reiterated the point,
while noting that $D_n\hbox{-}\sigma$ residuals were correlated with
mean surface brightness. Subsequently, J\o{}rgensen \etal\ (1993) used
a sample of Coma cluster early-type galaxies to show that
$D_n\hbox{-}\sigma$ is a nearly edge-on, but imperfect projection of
the fundamental plane relation.

More recently, J\o{}rgensen \etal\ (1996) showed that the fundamental
plane of early-type galaxies in Gunn $r$ follows the relation
\begin{equation}
r_e \propto \sigma^{1.24}\langle I\rangle_e^{-0.82}
\label{eq:fp}
\end{equation}
using 224 early-type galaxies in 11 clusters. Ellipticals and S0s
appear to follow the same relation, both with very low scatter. In
Coma, for example, the scatter is only 14\% in $r_e$ ({\it i.e.\/},
distance). In contrast to the Faber-Jackson relation which employed
elliptical galaxies as a family of ``standard candles,'' the tighter
fundamental plane and \dns\ scaling relations turn early-type galaxies
into accurate ``standard rods.''

Given the homogeneity of early-type galaxy stellar populations ({\it
e.g.\/}, Sandage \& Visvanathan 1978, Bower, Lucey, \& Ellis 1992) the
existence of a plane follows from basic principles (Faber \etal\
1987). While the details of the physical basis behind the empirical
relation remain elusive (Pahre 1998; though see Kelson 1998), the
general picture of early-type galaxies in dynamical equilibrium with
fairly homogeneous stellar populations seems to provide a reasonable
approximation to the observed fundamental plane (Kelson 1998, 1999).

With the assumption that early-type galaxies form a homologous family,
the combination of the virial theorem and the fundamental plane
implies that galaxy $M/L$ ratios are strongly correlated with their
structural parameters:
\begin{equation}
M/L \propto \sigma^{0.49} r_e^{0.22} \sim M^{1/4}
\label{eq:ml}
\end{equation}
Thus, when the \dns\ or fundamental plane relations are used as
distance indicators, one makes several important assumptions: (1)
$M/L$ ratios scale with galaxy structural parameters in the same way,
everywhere; and (2) early-type galaxies have similar stellar
populations (age, etc.), for a given galaxy mass, everywhere.

While these assumptions may not necessarily be valid, they have been
observationally verified by extensive work (Burstein \etal\ 1990,
J\o{}rgensen \etal\ 1996, though see Djorgovski \etal\ 1988). More
recently, however, Gibbons, Fruchter, \& Bothun (1998) have discovered
evidence that the fundamental plane scatter is related to environment,
such that clusters which are not X-ray luminous have poorly-defined
fundamental planes. However, such departures from universality are
relevant to the derivation of peculiar velocities, and are of limited
concern for the determination of the mean expansion rate of the local
universe. The chief requirement for our use of the fundamental plane
as a secondary distance indicator is that the slope and scatter of the
{\it mean\/} relation in the distant clusters be similar to the local,
calibrating samples. While this is shown explicitly in Section
\ref{fplocal}, we have an {\it a priori\/} expectation that this
assumption be valid for the following reasons: (1) the fundamental
plane, as an empirical relation, is a combination of a systematic
variation of stellar populations with galaxy mass with correlated
observational errors and selection biases (Faber \etal\ 1987, Kelson
1998, 1999); (2) this systematic trend in the properties of the
stellar populations is the same in Virgo and Coma (Bower \etal\ 1992);
and (3) residuals from the fundamental plane relation for cluster
E/S0s are uncorrelated with other stellar populations indicators ({\it
e.g.\/}, J\o{}rgensen \etal\ 1996, Kelson \etal\ 1999c).

Nevertheless, despite any uncertainties behind the origin and
universality of the fundamental plane, its utility as a distance
indicator has been empirically established, and consistently
reaffirmed ({\it e.g.\/}, J\o{}rgensen \etal\ 1996 and others). Since
their inception, several groups and teams of people have successfully
used the fundamental plane to measure distances to individual
elliptical galaxies, and to groups and clusters, in efforts to derive
values of the Hubble constant or map the local flow-field ({\it
e.g.\/}, Dressler \etal\ 1987, Wegner \etal\ 1996, Hudson \etal\
1997). Given the reliability with which this distance indicator can be
used, we now attempt to use the database of published Cepheid
distances to calibrate the fundamental plane, and derive a value for
$H_0$.


\section{The Data}
\label{data}

In this section, we discuss the sources of data used in our analysis,
derive structural parameters for a sample of early-type galaxies in
Leo I, Virgo and Fornax, collect useful data from the literature, and
discuss the typical uncertainties in our fundamental plane data. The
galaxies in the sample are listed in Table \ref{tab:params}.

The fundamental plane relies on two types of data: photometry and
spectroscopy. Galaxy imaging data, or aperture photometry, must be of
sufficiently high $S/N$ to derive effective radii ($\theta_e$, in
angular units; and $r_e$, in metric units) and mean surface
brightnesses within the effective radii. Furthermore, the photometric
calibration must be consistent for each set of galaxies to be placed
on the fundamental plane. Therefore, we choose to use the CCD
photometry of Tonry \etal\ (1997). As for the spectroscopy, the $S/N$
must be sufficient to measure internal kinematics from absorption line
widths. Our adopted velocity dispersions come from several reliable
sources, discussed below. Lastly, because our estimate of the Hubble
constant relies on direct comparisons of data from different sources,
considerable care was taken to measure the calibrators using the same
procedures which were used in analyzing the distant samples.

\subsection{Velocity Dispersions}

The Virgo and Fornax galaxies we use in this analysis were taken from
the Dressler \etal\ (1987), who reported velocity dispersions for 20
galaxies in Virgo and 8 in Fornax. These data were obtained as part of
a $B_T$-limited survey of elliptical galaxies, and the spectroscopy
for the Virgo and Fornax galaxies were obtained primarily at Las
Campanas Observatory (LCO). These Virgo and Fornax data form our
initial sample, but their numbers were subsequently reduced by two due
to a lack of precision photometry.

For the early-type galaxies in Leo I, we use Faber \etal\ (1989) who
report velocity dispersions for N3377 and N3379, corrected to an
effective aperture at the distance of Coma. We supplemented these with
central velocity dispersions for N3384 and N3412 taken from Fisher
(1997).

J\o{}rgensen \etal\ (1995a) tested the accuracy of the Dressler \etal\
(1987) dispersions. No significant systematic difference was found
between the dispersions measured by J\o{}rgensen \etal\ and those that
Dressler \etal\ obtained at LCO. The scatter between their
observations was \about 7\%, resulting in a random error in distance
for an individual galaxy of \about 9\%. Given this, the J\o{}rgensen
\etal\ (1995a) collection of spectroscopic parameters could be
homogenized and placed on a single consistent system in agreement with
Dressler \etal\ (1987).


\subsubsection{Velocity Dispersion Aperture Corrections}

Early-type galaxies have radial gradients in velocity dispersion,
$\sigma$, and radial velocity, $V_r$ ({\it e.g.\/}, Davies 1981, Tonry
1983). The implication is that the $\sigma$ one measures depends upon
the metric aperture used to extract the original spectra. Typically,
one's aperture is defined by an angular extent upon the sky, so the
metric aperture which defines the velocity dispersion typically
increases with distance to a given galaxy. This variation in aperture
with distance can be a source of systematic error for the fundamental
plane and must be accounted for when making distance determinations.

Early-type galaxies can be partially supported by rotation, and
velocity dispersion measurements are mixture of both pressure and
rotational support. The dependency of $\sigma$ upon aperture is not
straightforward to predict. The observation of a dispersion is akin to
taking a luminosity-weighted mean of the second moment of the
line-of-sight velocity distribution, $\sigma^2 + V_r^2$, within some
predefined spectrograph aperture. This second moment is also weighted
by the surface brightness distribution of the galaxy within the
aperture. Thus, measurements of $\sigma $ depends on each galaxy's
intrinsic distribution of orbits, as well as its light distribution
({\it e.g.\/}, Tonry 1983).

We chose to use the aperture correction prescription of J\o{}rgensen
\etal\ (1995a). They used photometry and long-slit spectroscopy from
the literature to construct dynamical models of early-type galaxies.
These models were projected onto the sky, and ``observed'' in order to
determine how the measured velocity dispersion depends upon aperture
size. They found that the measured dispersions described a power law
function of the aperture size, $d$,
\begin{equation}
\log \sigma_{\rm cor} =
\log \sigma_{\rm obs} + 0.04 \times (\log D_{\rm cor} - \log D_{\rm obs})
\label{eq:apcor}
\end{equation}
where $D_{\rm cor}$ is the nominal aperture, and $D_{\rm obs}$ is the
aperture of the observation. This prescription implicitly incorporates
both the declining velocity dispersion profiles of early-type
galaxies, and the rising rotation curve as well.

We can use Eq. \ref{eq:apcor} to correct the observed velocity
dispersions of galaxies in Leo I, Virgo, and Fornax to equivalent
measurements from a nominal aperture with a $3 \Sec 4$ diameter at the
distance of Coma, the system used for the J\o{}rgensen \etal\ (1996)
sample. This aperture is equivalent to a diameter of $D\approx 31''$ at
the distance of Leo I, $\about 20''$ at the distance of Virgo, and
$\about 15''$ for Fornax. The Dressler \etal\ (1987) Virgo and Fornax
measurements were obtained using apertures of $D=16''$ so the aperture
corrections for the Dressler \etal\ dispersions are small ($-1\%$ in
Virgo, $+1\%$ in Fornax). The Fisher (1997) measurements for N3384 and
N3412 were derived using an aperture of $4'' \times 2''$, equivalent to
a circular aperture of $D=3\Sec 8$, and thus required an aperture
correction of $-8\%$. The Faber \etal\ (1989) data for N3377 and N3379
had been corrected for aperture, by $-5\%$, but we removed their
correction in order to apply the J\o{}rgensen \etal\ (1995a)
prescription's aperture correction of $-9\%$.


\subsection{Photometric Structural Parameters}

In this section, we discuss the determination of the photometric
parameters required for the analysis of the fundamental plane. In
particular, we derive the effective radii and surface brightnesses for
the galaxies in the three nearby clusters, and discuss the
transformation of the photometry to Gunn $r$, the system used by
J\o{}rgensen \etal\ (1996) for the sample of distant clusters.

\subsubsection{Effective Radii and Mean Effective Surface Brightnesses}

Tonry \etal\ (1997) observed more than 150 galaxies as part of a program
to measure surface brightness fluctuations (SBF) in a large sample of
early-type galaxies and spiral bulges. These SBF measurements provide a
direct estimate of the distance to each galaxy and these distances are
being used to determine $H_0$ and map the nearby peculiar velocity field
({\it e.g.\/}, Tonry \etal\ 1997). Those authors used CCD imaging in two
colors, Johnson $V$ and Cousins $I_c$, from several cameras and
telescopes, making a painstaking attempt to unify the calibrations in
one homogeneous system. Multiple observations of each galaxy showed that
the $V$-band CCD photometry has uncertainties of $\pm 0.02$ mag, and the
$V-I_c$ colors have uncertainties of $\pm 0.02$ mag (see \S \ref{errors}
for further discussion). These data prove to be ideal for fundamental
plane analysis.

\begin{figure*}
\centerline{\mkfigbox{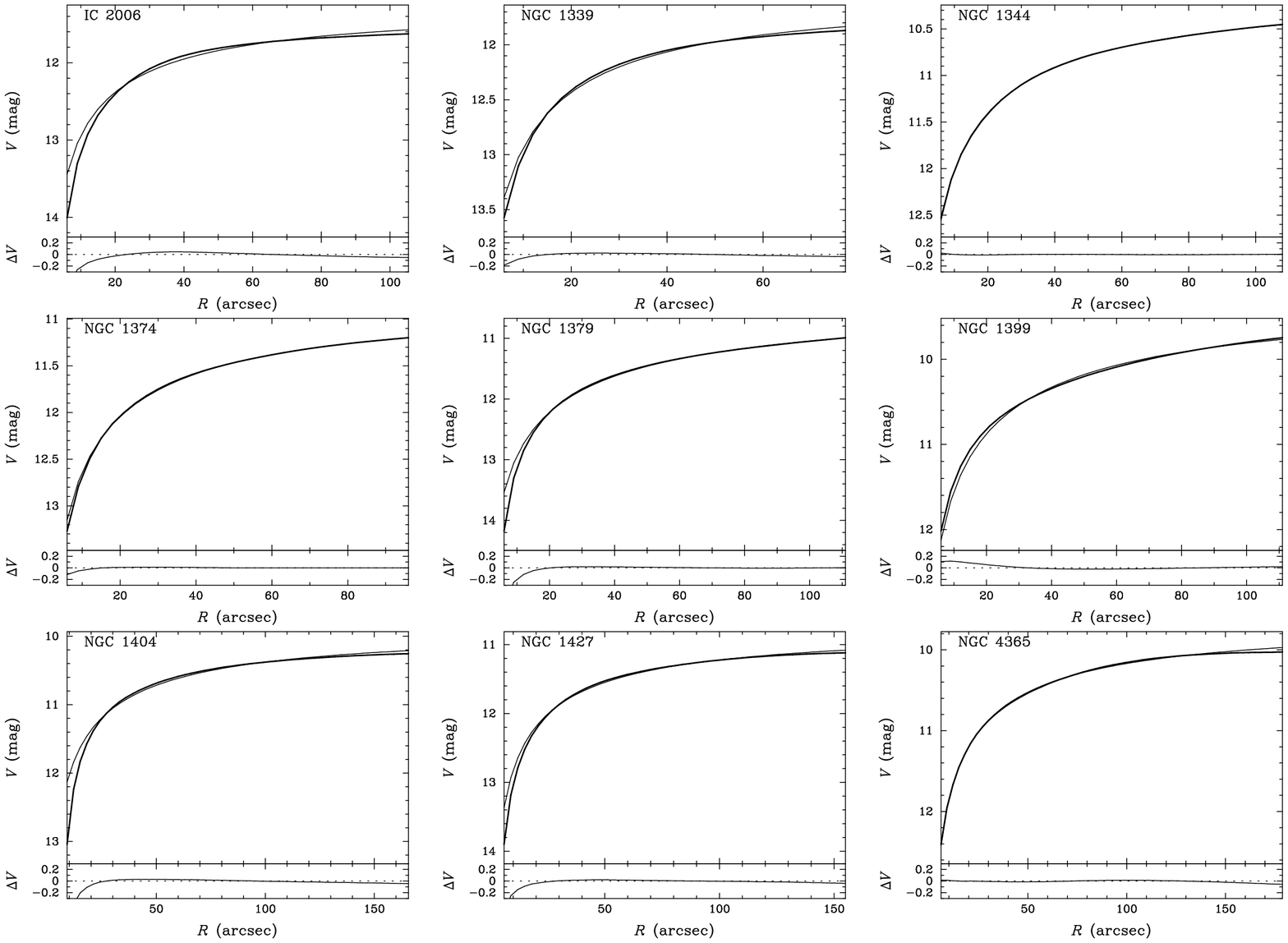}{0.8\textwidth}}
\figcaption{
The Johnson $V$ circular aperture growth curves of Tonry \etal\ (1997)
are shown as thick solid lines, while the integrated $r^{1/4}$-law
profiles are shown as thin solid lines. In the lower panels, the
residuals from the $r^{1/4}$-law fit are shown.
\label{fig:growth}}
\end{figure*}

The Tonry \etal\ (1997) sample included circular aperture photometry on
18 Virgo and eight Fornax galaxies of the ones listed in Dressler \etal\
(1987), and four early-types in Leo I. Tonry \etal\ masked out bright
stars, overlapping objects, saturated pixels, other image defects, and
removed the sky background. The $S/N$ is extremely high, with negligibly
small formal errors. The concentric aperture photometry profiles in the
$V$ band are shown as thick solid lines in Figure \ref{fig:growth}.

These data are well-suited for the fitting of parameterized growth
curves. For consistency with the J\o{}rgensen \etal\ (1996)
fundamental plane we fit integrated $r^{1/4}$-law profiles to the
galaxy growth curves. The curves were iteratively fit to a limiting
radius of 2-3 times $r_e$ in an effort to minimize the effects of
sky-subtraction errors in the fit. In Figure \ref{fig:growth}, we show
the fitted integrated $r^{1/4}$-law by a thin solid line. Residuals
from this fit are shown as well. The resulting $V$-band effective
radii ($\theta_e$, in arcsec) and $\langle I\rangle_e$ are listed in
Table \ref{tab:params}.

The typical uncertainty in the fitted $\theta_e$ is 9\%. However,
there are some galaxies for which the errors are as large as 20-30\%.
Because the errors in $\theta_e$ and $\langle I \rangle_e$ are
strongly correlated (see \S \ref{dnsigma}, and, {\it e.g.\/}, Saglia
\etal\ 1997, Kelson \etal\ 1999b), the uncertainties in the surface
brightnesses show correspondingly large, correlated errors (with a
correlation coefficient of $-0.73$; also see Kelson \etal\ 1999b).
What is important to note, however, is that the error in the
combination of $\theta_e$ and $\langle I\rangle_e$ that enters the
fundamental plane ($\theta_e \langle I\rangle_e^{0.82}$, the
fundamental plane parameter) will be much smaller than the errors in
the individual structural parameters (typically on the order of $\pm
3\%$).

The tabulated errors in $\theta_e$ and $\langle\mu\rangle_e$ do not
include systematic errors which are incurred by adopting a parameterized
profile (the de Vaucouleurs profile). Therefore, we experimented with
generalizing the surface brightness profiles to the Sersic (1966)
$r^{1/n}$-law. By generalizing the growth curves we find a mean
systematic offset of $-1\%$ in distance in the combination of $\theta_e$
and $\langle I\rangle_e$ that enters the fundamental plane. The
1-$\sigma$ scatter about this mean offset is $\pm 5\%$. We conclude that
any systematic effect due to galaxy profile shapes has no significant
impact on our results. We are confident that no significant systematic
errors are incurred by adopting the de Vaucouleurs $r^{1/4}$-law as the
parameterization of the galaxy surface brightness profiles (see Kelson
\etal\ 1999 for a more detailed discussion).

To summarize, we anticipate that the systematic errors in the
fundamental plane parameters are $< 1\%$. The random errors dominate for
a given galaxy and are $\pm 5\%$ in distance. These uncertainties do not
include, for example, the systematic errors of the photometric
calibration, which will be discussed later.


\subsubsection{Transformation to Gunn $r$ Photometry}
\label{transform}

The J\o{}rgensen \etal\ (1996) sample of 224 early-type galaxies in 11
nearby clusters was based on a compilation of photometry in Gunn $r$
(J\o{}rgensen \etal\ 1995b). Thus, in order to directly compare the
fundamental plane relations in Leo I, Virgo, and Fornax with the
fundamental plane relations of J\o{}rgensen \etal, we need to
transform our $V$-band surface brightnesses to $r$. We employ the
photometric transformation of J\o{}rgensen (1994), who used 75
photometric standards to derive the following transformation between
$V$, $(B-V)$ to Gunn $r$:
\begin{equation}
r = V + 0.273 - 0.486 (B-V)
\label{eq:xfm}
\end{equation}
The Tonry \etal\ (1997) photometry is in $V$ and $I_c$, so we used
Frei \& Gunn \etal\ (1994) to derive $(B-V) \approx 0.814\times
(V-I_c)$. Before applying the transformation, the Galactic foreground
extinctions for each galaxy are individually removed. The extinction
estimates, given in Table \ref{tab:params}, were derived from Schlegel
\etal\ (1998). The reddening law of Cardelli, Clayton \& Mathis (1989)
was used to calculate the extinctions in the $V$ and $I_c$ passbands.

Small corrections due to $(1+z)^4$ surface brightness dimming, and small
$K$-corrections were applied to the Tonry \etal\ (1997) photometry as
well. In Gunn $r$, the $K$-correction is well described by
$2.5\log(1+z)$ (J\o{}rgensen \etal\ 1995b). The published J\o{}rgensen
\etal\ data were already corrected for these effects.

We defer discussion of the errors in photometry and photometric
transformation to Section \ref{errors}.


\subsubsection{Derivation of $r_n$ in the Gunn $r$ Passband}
\label{findrn}

The fundamental plane and \dns\ relations are closely related (Faber
\etal\ 1987, J\o{}rgensen \etal\ 1993), and so we derive the isophotal
radii of the Leo I, Virgo, and Fornax galaxies to facilitate comparison
of the two distance indicators. Similar to the effective radii derived
earlier, we define $\theta_n$ to be the angular radius within which each
galaxy has a specific mean surface brightness, and $r_n$ is the
equivalent radius expressed in metric units.

Using the transformation of Equation \ref{eq:xfm}, and the mean
$(V-I_c)$ colors of the galaxies, we transformed the concentric
circular photometry to Gunn $r$ in order to derive isophotal radii
$\theta_n \equiv D_n/2$ in the same filter as was used to derive
$\theta_n$ for the Coma galaxies. J\o{}rgensen \etal\ (1995b) defined
$\theta_n$ in Gunn $r$ as the radius within which the mean surface
brightness is 19.6 mag/arcsec$^2$. The values of $\theta_n$ are also
listed in Table \ref{tab:params}.

Our measurements of $\theta_n$ should be as well-determined as the
fundamental plane parameters (see previous discussion). Errors in
$\theta_n$ come from uncertainties in the transformation to Gunn $r$,
the calibration of the original $V$ photometry, uncertainties in the
color gradients, photon statistics, and the noise characteristics of
the detector. Because the color gradients in these galaxies are
typically small ($|d(V-I_c)/d\log \theta| \simlt 0.05$ mag/dex), we do
not expect them to significant affect the $r_n$ measurements (also see
\S \ref{errgrad}). For the moment, we assume that the errors in
$\theta_n$ are similar to the errors in $\theta_e\langle
I\rangle_e^{0.82}$.


\section{The Cepheid Distances to Leo I, Virgo, and Fornax}
\label{adopt}

Cepheids remain a key component of the distance scale ladder for
individual galaxies, and hence for the calibration of such secondary
distance indicators. One assumption in our analysis is that the Cepheid
distances to the spiral galaxies in these clusters/groups are
appropriate for the early-types which we have also associated with those
clusters/groups. A good estimate for the error incurred by this
assumption is not easy to determine, but the low scatter among the
Cepheid distances to Virgo may indicate that this error is small. This
uncertainty is one of many sources of error to be discussed in the
discussion of the error budget in \S \ref{errors}. The published Cepheid
distances to the spiral galaxies in the Leo I group and the Virgo and
Fornax clusters are listed in Table \ref{tab:calibs}, along with their
references (see Ferrarese \etal\ 1999). In deriving a weighted mean
distance to Virgo, we exclude NGC 4639.

Our adopted distance mean moduli to Leo I, Virgo, and Fornax, weighted
by the random errors, are $30.08 \pm 0.11 \pm 0.16$ mag, $31.03\pm 0.04
\pm 0.16$ mag, and $31.60\pm 0.14 \pm 0.16$ mag, respectively. These
distance moduli correspond to metric distances of $10.4\pm 0.6\pm 0.8$
Mpc, $16.1\pm 0.3 \pm 1.2 $ Mpc, and $20.9\pm 1.4\pm 1.6$ Mpc. These
distances are tied to the LMC distance modulus, where we have adopted
$\mu_{\rm LMC}=18.50 \pm 0.13$ mag.

The uncertainties in our distances to Leo I, Virgo, and Fornax are the
systematic and random errors, respectively, arising from the Cepheid
distance estimates alone. For the total random uncertainties, we have
added the random errors in the individual Cepheid distances in
quadrature with the standard error of the mean distance. The various
sources of error in the Cepheid distances are listed in Table
\ref{tab:errors}. Some of these components are exacerbated by the
de-reddening procedure, such as the uncorrelated errors in the F555W and
F814W photometric zero points (which systematically impacts $H_0$), or
the errors in the F555W and F814W aperture corrections (which is a
random error for each Cepheid-based distance). Thus, the systematic
errors given in the previous paragraph are simply a combination of
uncertainty in the distance to the LMC, the uncertainties in the LMC PL
relations themselves, and the uncertainties in the WFPC2 photometric
calibration. The estimates for the random errors to each cluster are
assumed to be comparable to the random error to a given galaxy, divided
by root-$N$.

After some of these distances were published, Kennicutt \etal\ (1998)
measured a marginally significant dependence of the Cepheid PL relation
on metallicity of $\Delta \mu_{VI_c}=-0.24\pm 0.16$ mag per dex in [O/H].
Differences between the abundances in our Cepheid fields and the LMC
lead to systematic errors in the inferred distances. For some secondary
indicators, the Cepheid calibrating fields span a wide range of oxygen
abundances with a mean abundance equal to that in the LMC. Such is the
case for the Tully-Fisher relation, for which the effects of metallicity
incur a net effect on the Hubble constant of zero (Sakai \etal\ 1999).
In this paper, the number of Cepheid calibrators is small, with a mean
metallicity greater than that in the LMC. Thus metallicity corrections
to the PL relation will have a systematic effect on $H_0$ and
this systematic effect will be discussed in \S \ref{syscor}.


\section{The Fundamental Plane and a Determination of the Hubble
Constant}
\label{fplocal}

\subsection{The Fundamental Plane Relations of Leo I, Virgo, and Fornax}
\label{gammas}

\begin{figure*}
\centerline{ \mkfigbox{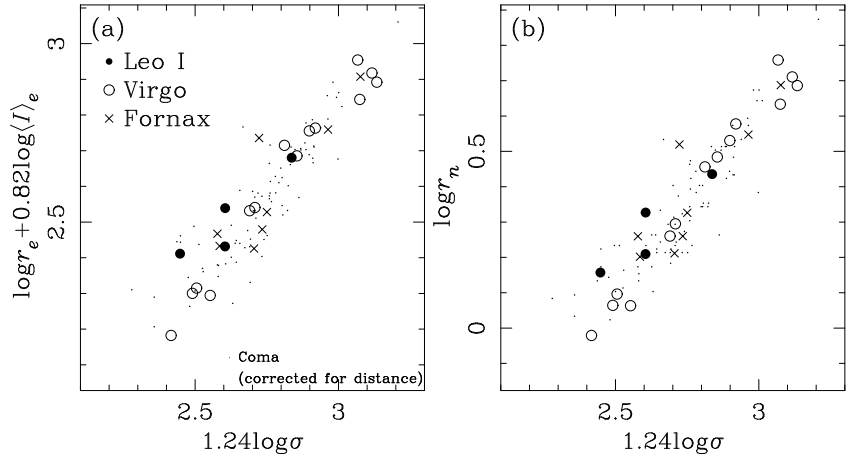}{0.6\textwidth}}
\figcaption{
The (a) fundamental plane and (b) \dns\ relations in Leo I, Virgo, and
Fornax, where distance effects have been removed. The Coma sample from
J\o{}rgensen \etal\ (1995ab) is shown, corrected for distance, by
small points.
\label{fig:fplane}}
\end{figure*}

The fundamental plane in the three clusters is shown in Figure
\ref{fig:fplane}(a). For our analysis, we convert the surface
brightnesses to units of $L_\odot/$pc$^2$ in the bandpass of interest
(Gunn $r$), and refer to the effective radii by $r_e$, in metric units
of kpc, using the Cepheid distances to each cluster or group adopted
in the previous section.

Fixing the slope of the fundamental plane to that found in the distant
cluster sample by J\o{}rgensen \etal\ (1996), the zero point is defined
as $\gamma \equiv \log r_e -1.24 \log \sigma + 0.82 \log \langle
I\rangle_e $. In J\o{}rgensen \etal\ (1996), the sample sizes were
large, and as a result, those authors defined the zero points using the
median, a technique that is robust to outliers. However, because our
samples in Leo I, Virgo and Fornax are so small, we opt for the mean
zero point, rather than the median, even though using the median does
not lead to significantly different results.

Because the values of $r_e$ for the Leo I, Virgo, and Fornax galaxies
have all been expressed in metric units, the galaxies should all lie
along the same line in the figure. Any scatter in the figure from
galaxy to galaxy, or from cluster to cluster should only arise from
(1) an error in Cepheid distance; (2) a breakdown in the assumption
that that Cepheid distance applies to the early-types in the group or
cluster; (3) errors in the fundamental plane parameters themselves; or
(4) inhomogeneities in galaxy $M/L$ ratio from galaxy to galaxy. These
potential sources of error will be discussed below in \S \ref{errors},
but they are clearly not large given that the three cluster samples
agree so well in the figure.

Before deriving the fundamental plane zero point, we note that the CCD
imaging of N4552 was saturated in the core. Thus, its derived mean
surface brightness is in error, and we exclude it from subsequent
analysis. Furthermore, N4489 is 0.09 mag bluer ($3\sigma$) than the mean
$(V-I_c)$ color of the other early-type galaxies and is also excluded.
Lastly, Tonry \etal\ (1997), using SBF measurements, argue that N4365
and N4660 are not {\it bona fide\/} members of Virgo. As a result of
removing these objects, our final sample is comprised of 4 galaxies in
Leo I, 14 in Virgo, and 8 in Fornax, as shown in the plot of the
calibrated fundamental plane in Figure \ref{fig:fplane}(a).

In Table \ref{tab:mzpts}, we give mean fundamental plane zero points,
$\langle \gamma \rangle$, derived in each cluster and their formal
errors. The external systematic and random errors are shown in the last
column. Given the magnitude of the internal and external random errors,
the zero points in the three clusters agree remarkably well. Taking a
weighted average of the three gives
$\langle\langle\gamma\rangle\rangle=-0.173\pm 0.013$. By computing the
mean zero point using all 26 galaxies at once, instead of averaging the
individual zero points of the clusters, one finds no significant change
in the mean fundamental plane zero point,
$\langle\langle\gamma\rangle\rangle$. This error is the quadrature sum
of the standard error of the mean and the random errors in the Cepheid
distances to the clusters. The scatter among the three cluster zero
points is consistent with the uncertainties in the individual zero
points and the random errors in the Cepheid distances, suggesting that
the intrinsic differences in the mean galaxy $M/L$ ratios among the
three clusters are not large. None of the individual cluster zero points
differ from the mean by more than 1.5$\sigma$, when one takes into
consideration the internal errors, and the random errors in the Cepheid
distances. The full error budget will be discussed below in \S
\ref{errors}.

This reported zero point is only valid for the set of fundamental
plane slopes adopted above. Changing $\alpha$ and $\beta$ will produce
different zero points in both the calibrating and distant samples,
leading to small changes in the derived relative distance. We will
explore this source of uncertainty in \S \ref{errors}.


\subsection{The Fundamental Plane in Distant Clusters and the Hubble
Constant}

The fundamental plane zero point explicitly relates angular size of
$\theta_e$ to a metric scale. By comparing the zero point of the
metric ($\log r_e$) fundamental plane to the angular ($\log \theta_e$)
fundamental plane in a distant cluster, one directly infers its
distance.

For example, J\o{}rgensen \etal\ (1995ab) provide fundamental plane
data for 81 galaxies in Coma. Using $\theta_e$ in radians, one finds a
zero point for Coma of $\gamma_{\rm C} = -5.129\pm 0.009$. Thus the
distance to Coma, in Mpc can be written as $\log d=5.129+\langle
\gamma\rangle-3$. (The additional constant converts the distances to
units of Mpc.) Using the weighted mean zero point defined in the
previous section, one derives a distance to Coma of $90\pm 6$ Mpc.
Correcting for this distance, we show the galaxies in the Coma sample
in Figure \ref{fig:fplane}(a) as small points. Note that the nearby
galaxies follow the same relation as that in Coma, validating our
assumption that the distant fundamental plane be of the same shape and
scatter as that in the nearby clusters. Using a recession velocity for
Coma of 7143 \kms, one obtains $H_0=79 \pm 6$ \kms\Mpc-1. These
include only the random errors, all of which are discussed with the
total error budget in \S \ref{errors}.

Distant clusters can have peculiar velocities, which may lead one to
deduce incorrect values for the Hubble constant. Therefore, we choose to
derive a mean value of the Hubble constant using several distant
clusters, with the hope that errors in individual measurements due to
the peculiar velocity field will be reduced (see below), by $\sqrt {N}$.
For this paper, we use the 11 distant clusters of J\o{}rgensen \etal\
(1996).

In Table \ref{tab:allh0}, we list all 11 clusters, the number of
galaxies used in each of them, the fundamental plane zero points for
each cluster (again, using $\theta_e$ in radians), the \rms\ scatter,
the implied distance to each cluster, the recession velocity of each
cluster with respect to the cosmic microwave background (CMB) and the
implied Hubble constant, and the recession velocity of each cluster
with respect to the flow-field model of Mould \etal\ (1999a) and the
implied Hubble constant. The errors which are listed are the
quadrature sum of the random errors in the Cepheid-based zero point,
the random errors in the zero points of each cluster, and the random
errors due to uncertainties in the fundamental plane slope (see \S
\ref{sloperr}).

The average value of $H_0$, weighting by the internal errors, over all
11 clusters is $\langle H_0\rangle =82\pm 5$ \kms\Mpc-1. This value is
produced whether we use the CMB or flow-field recession velocities.
Several of the distant clusters have poorly defined fundamental plane
zero points. The largest individual zero point uncertainty is in Grm
15, for which the median zero point leads to a 15\% larger value for
$H_0$. Averaging over the 11 clusters and using the median zero points
leads to a net increase of 3\% in the Hubble constant. When using the
five clusters with samples of $N\ge 20$ galaxies, we find a 2.5\%
decrease in the Hubble constant. We conclude that our results are
insensitive to the distant cluster sample sizes.

For the five clusters with good statistics ($N\ge 20$ galaxies), the
weighted \rms\ scatter in the individual values of $H_0$ is 4\%. We
interpret this scatter as the uncertainty due to the peculiar velocity
field when using a single distant cluster and incorporate this
uncertainty into the error budget. Averaged over the 11 clusters, the
resulting uncertainty in the Hubble constant due to the peculiar
velocity field is expected to be $1\%$. To help test the effect of the
velocity field, we computed $H_0$ using the six clusters with $cz_{\rm
CMB}\ge 5000$ \kms, and the four clusters beyond $cz_{\rm CMB}\ge
7000$ \kms. A net difference of 1\% in $H_0$ is observed. While the
peculiar velocities for the distant clusters do contribute small
random errors, we conclude that they are not a dominant source of our
uncertainties in $H_0$. If one uses the flow-field velocities of Mould
\etal\ (1999a), the results differ by no more than 1\%.

The errors in the values of $H_0$ given above are the quadrature sum
of the random error in the Cepheid-based zero point, the formal errors
in individual zero points of the distant clusters, the random errors
due to the uncertainties in FP slopes, and the error in the unknown
peculiar velocity field.

\medskip

Before correcting the value of $H_0$ for systematic effects, we now
discuss a determination of $H_0$ using the \dns\ relation. Following
the analysis, we discuss \dns\ in the context of the fundamental
plane, and show why the value of the Hubble constant derived using the
fundamental plane is to be preferred.


\section{The \dns\ Relation}
\label{dnsigma}

The \dns\ relation has traditionally been thought of as equivalent to
the fundamental plane (Faber \etal\ 1987 and Djorgovski \& Davis
1987). However, J\o{}rgensen \etal\ (1993) argued that the two
relations are not equivalent, that \dns\ is not quite identical to the
edge-on projection of the fundamental plane. Nevertheless, the \dns\
relation can provide an additional test of the robustness of our
results and illuminate our sensitivity to the systematic differences
between the two scaling relations.

\subsection{The \dns\ Relations of Leo I, Virgo, and Fornax}

In parallel with the fundamental plane analysis, though using the
isophotal radii, $r_n$, derived in \S \ref{findrn}, we define the \dns\
zero points as $\delta\equiv \log r_n - 1.24 \log \sigma$, much like the
fundamental plane zero points were defined above. In Figure
\ref{fig:fplane}(b), we plot the \dns\ relations of Leo I, Virgo, and
Fornax. For this figure, the $r_n$ values are expressed in kpc.

In Table \ref{tab:mzpts}, we list the values of $\langle \delta\rangle$
for each group/cluster. The \dns\ zero points also agree remarkably well
with each other, which should not be surprising given the intimate
connection between \dns\ and the fundamental plane. A weighted mean
gives a value of $\langle\langle\delta\rangle\rangle=-2.395\pm 0.013$.
This error is the quadrature sum of the standard error of the mean and
the random errors in the Cepheid distances to the clusters.

\subsection{The \dns\ Relation in Distant Clusters}

We can now compare this calibrated \dns\ zero point with the \dns\
relations of the clusters in J\o{}rgensen \etal\ (1996) to measure their
distances, and thus derive a value of the Hubble constant for comparison
with the one given previously in \S \ref{fplocal}.

The results for each cluster are given in Table \ref{tab:allh0}. For
example, the 81 galaxies in Coma have a mean \dns\ zero point of
$\delta_{\rm C} = -7.373\pm 0.009$, in which $\theta_n$ is expressed
in radians, Thus the distance to Coma, in Mpc can be written as $\log
d=7.373+\langle \delta\rangle-3$. Using the weighted mean of the Leo
I, Virgo, and Fornax zero points, one derives a distance to Coma of
$95\pm 7$ Mpc. The \dns\ relation for Coma is shown in Figure
\ref{fig:fplane}(b) by the small points. The implied value of the
Hubble constant is $H_0=75 \pm 5$ \kms\Mpc-1. These errors, too, are
the quadrature sum of the random errors.

Averaging over the 11 clusters, one finds $\langle H_0\rangle=79\pm 6$
\kms\Mpc-1, a difference of only 4\% compared to the value found using
the fundamental plane.


\subsection{The Relation between \dns\ and the Fundamental Plane}
\label{dnfprel}

The assumption that the two relations are equivalent implies that $r_n
\equiv r_e\langle I\rangle_e^{0.82}$. Using the Tonry \etal\ (1997)
photometry we can investigate at what point this assumption breaks
down. Dressler \etal\ (1987) showed that for a set of galaxies with
identical growth curves, the ratio of $r_n/r_e$ is indeed a power-law
function of $\langle I\rangle_e$ (without specifying the exponent). By
fitting a universal growth curve to any set of galaxies with a range
of profile shapes, the exponent is indeed constant, and furthermore is
only a function of the what growth curve is used in the fitting
(Kelson \etal\ 1999b).

\begin{figure*}
\centerline{ \mkfigbox{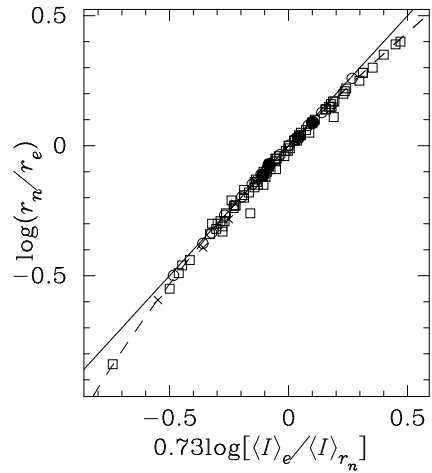}{0.3\textwidth}}
\figcaption{
The relationship between $r_n$ and $r_e$, $\langle I\rangle_e$. The
symbols are as in Figure \ref{fig:fplane} with the addition of the Coma
sample as squares. The solid line has a slope of unity, representing the
leasts-squares fit to the data (see \S \ref{dnfprel}). The dashed line
follows a pure $r^{1/4}$-law growth curve. Thus, the \dns\ relation is
a combination of the fundamental plane and the error correlation
between $r_e$ and $\langle I\rangle_e$.
\label{fig:dnfp}}
\end{figure*}

We therefore performed a simple test to determine the relationship
between $r_n$, $r_e$ and $\langle I\rangle_e$. For the four Leo I
galaxies, the 14 Virgo galaxies and the eight Fornax galaxies, plus
the J\o{}rgensen \etal\ (1996) sample of 81 Coma galaxies, we find
\begin{equation}
\log {(r_n/r_e)} = (0.73\pm 0.01)\times \log
[\langle I\rangle_e/ \langle I\rangle_{r_n}]
\end{equation}
with a scatter of about 0.01 dex. This relation holds in each of the
four clusters, as shown in Figure \ref{fig:dnfp}. The solid line, with
the slope of unity in this projection, represents the above relation.

The above expression relates $r_n$ to effective radius and mean
surface brightness and explicitly shows that the \dns\ scaling
relation has a different shape from the fundamental plane. The \dns\
relation can be thought of as a view of the fundamental plane which
has been skewed by the correlated errors between $r_e$ and $\langle
I\rangle_e$. Thus we have confirmed the result of J\o{}rgensen \etal\
(1993), who concluded that \dns\ is a slightly curved projection of
the fundamental plane.

The curvature of this projection is expected from the fact that the
surface brightness profiles of early-type galaxies, in general, are
well-described by $r^{1/4}$-laws. In Figure \ref{fig:dnfp}, the dashed
line shows the relation between radius and mean surface brightness
within that radius for an $r^{1/4}$-law. The curve is not a fit to the
data but is simply a projection of the $r^{1/4}$-law growth curve. The
conclusion is that $r_n$ is not a linear approximation to the
fundamental plane parameter, and that the accuracy of \dns\ as a
distance indicator is related to the range of galaxy sizes used in the
scaling relation.

Both \dns\ and the fundamental plane are affected by the correlation of
error in $r_e$ and $\langle I\rangle_e$. For the fundamental plane,
galaxy scale lengths are difficult to measure to an accuracy of better
than 50\% (Caon \etal\ 1993, Kelson \etal\ 1999b). Monte Carlo
simulations of the fundamental plane indicate that such large errors in
half-light radii do not increase the scatter in the fundamental plane
because the underlying scaling relation is reasonably parallel to the
error correlation vector (within $\simlt 15^\circ$, Kelson 1999). In
other words, large errors in $r_e$ do not necessarily contribute to
large errors in distance. In the case of the fundamental plane, the
coefficient on the surface brightness term is not equal to
$\beta=-0.73$, as it is implicitly in \dns. Therefore, we conclude that
the FP is likely to be less-biased by the error correlation, and will be
a more reliable distance indicator.


\section{The Scatter in the Fundamental Plane and \dns\ relations}
\label{scatter}

The observed scatter in the Virgo fundamental plane and \dns\ relations
is very low; the 1-$\sigma$ width is $\pm 0.045$ dex in distance,
implying a $\pm 10\%$ uncertainty in the distance to any individual
galaxy. This low scatter is remarkable given that the uncertainties in
the velocity dispersion have been estimated at $\pm 9\%$ for a single
galaxy, and that the errors on our fundamental plane parameters are
probably no larger than $\pm 5\%$. Furthermore, the Virgo galaxies have
spatial extent on the sky, and there must be some corresponding depth to
the sample. Our sample has an approximate \rms\ scatter on the sky
equivalent to $3^\circ$, which, for a spherical distribution, implies an
additional $\pm 5\%$ scatter in distance along the line of sight. There
is very little room left for other sources of scatter and it is likely
that the estimates on the uncertainties of $\sigma$ and \fpparam{-0.82}
have been overestimated. With all of these sources of scatter, we
conclude that the intrinsic scatter in the fundamental plane of Virgo
must be very low, indicating remarkable homogeneity among their stellar
populations.

If environment does play a key role in determining galactic $M/L$
ratios, then the fundamental plane ellipticals in Virgo must be very
similar indeed to the inner regions of Coma, where the observed scatter
is also extremely low ($\pm 14\%$ in distance for a single galaxy). 

The 1-$\sigma$ scatter in the Fornax sample, is $\pm 0.092$ dex,
equivalent to 21\% in distance (also seen by D'Onofrio \etal\ 1997).
There is no obvious reason for this increase. The Fornax galaxies do not
display any obvious morphological peculiarities, nor are their internal
kinematics significantly different from normal ellipticals (Bender,
Saglia, \& Gerhard 1994). NGC 1344 has the largest residual and by
excluding it from the zero point determination, the remaining seven show
a scatter of $\pm 0.060$ dex. This scatter implies that these Fornax
ellipticals have an extra 10\%, in distance, added in quadrature with
the Virgo scatter, which would suggest (1) that the cluster is quite
elongated along the line of sight; or (2) that the remaining seven
early-type galaxies have slightly more scatter in their stellar
populations. However, there is no obvious {\it a priori\/} reason to
exclude NGC 1344 from the analysis, and doing so leads to no significant
change in the Hubble constant, given the weight of the other seven
galaxies in the cluster, and the weight that the Fornax zero point is
given compared to Virgo or Leo I.

Given the range of Cepheid distances ($0.4$ mag), perhaps the
ellipticals span a large range as well; however, assuming the same
internal errors in the Virgo and Fornax data, the Fornax plane
apparently has an additional 18\% scatter added in quadrature. Only if
Fornax is the chance superposition of two groups of ellipticals (near,
with N1326A and N1365, and far, with N1425), can this be achieved. There
is no strong evidence for this, but it is consistent with the suggestion
by Suntzeff \etal\ (1999) that the hosts of the Type Ia supernovae in
the direction of Fornax may be lying at a distance of 21.6 Mpc.

While the increased scatter in Fornax leaves open questions about the
uniformity of early-type stellar populations and the intrinsic spread in
the distances to Fornax ellipticals, we note that these uncertainties
have a small impact on our value of the Hubble constant, because we use
Leo I and Virgo as well, with Virgo having the highest weight. The
increased weight that has been given to Virgo has its own implications,
and we discuss those further below, in \S \ref{gonz}.

\medskip

In the following sections we address the systematic corrections to the
derived value of $H_0$, and potential sources of error in the current
analysis.


\section{Systematic Corrections to the Hubble Constant}
\label{syscor}

In this section we discuss three potential sources of systematic
error, and the corrections we make to our determination of $H_0$ for
these systematic effects. The first systematic correction is due to
the uncertain metallicity correction to the Cepheid P-L relation.
Next, the cluster population incompleteness bias is discussed in the
context of the fundamental plane samples. Lastly, we address the
systematic correction to $H_0$ which arises by adopting distances to
Leo I, Virgo, and Fornax which are derived from Cepheid distances to
spirals selected from an extended distribution along the
line-of-sight.

\subsection{The Metallicity Correction to the Cepheid Distances}

Kennicutt \etal\ (1998) found that the Cepheid P-L relation has a
small dependence upon [O/H] abundance, such that two-color $VI_c$
distance determinations are in error by $\Delta \mu_{VI_c} = -0.24\pm
0.16$ mag/dex. The trend is such that metal-rich Cepheids appear
brighter and closer than metal-poor ones. Because this determination
is not very significant, it has not been factored into our
determination of $H_0$.

Nevertheless, the uncertain metallicity correction is a source of
systematic error, and we now estimate to what extent our adopted
distances to Leo I, Virgo, and Fornax, are affected. The metallicity
corrected distance moduli increase $8\%$, $6\%$ and $4\%$ for Leo I,
Virgo, and Fornax, respectively, to $30.27 \pm 0.11 \pm 0.16$ mag,
$31.17 \pm 0.05 \pm 0.16$ mag, $31.68 \pm 0.16 \pm 0.16$ mag. These
correspond to metric distances of $11.3\pm 0.6\pm 0.9$ Mpc, $17.1\pm
0.4 \pm 1.3 $ Mpc, and $21.7\pm 1.7\pm 1.7$ Mpc, respectively.

Correcting for metallicity increases all of the adopted distances, and
thus leads to a decrease in $H_0$. Using the metallicity correction
derived by Kennicutt \etal\ (1998) leads to a value of $H_0=77\pm 5$
\kms\Mpc-1. Therefore, if one were to use of the Kennicutt \etal\
metallicity correction would decrease the Hubble constant by $6\%\pm
4\%$.


\subsection{Cluster Incompleteness Bias}
\label{cpib}

The ``cluster population incompleteness bias'' ({\it e.g.\/},
Federspiel \etal\ 1994) occurs because one is attempting to measure
the distance to standard candles in a magnitude-limited sample. One
can simulate the magnitude of the bias by sub-sampling the Coma
galaxies. We randomly select subsets of the Coma sample, imposing
selection criteria in $r_e$, or in apparent magnitude, such that the
randomly selected subset of the Coma data has the same size and depth
as the samples in each of the 11 clusters. By performing 500
iterations of this selection for each cluster, we can estimate the
systematic biases in the individual cluster zero points. In doing
these Monte Carlo experiments, we find that the systematic bias is
less than 1\%.

The effect can be tested using the data as well, but the results are
inconclusive because of small number statistics in most of the
clusters. J\o{}rgensen \etal\ (1996) showed that their samples are
reasonably complete down to $\sigma > 100$ \kms. By including only
those galaxies in the distant clusters with $\sigma > 100$ \kms, the
mean Hubble constant, found by minimizing the square of the residuals
in each of the clusters, decreases by 2\%. Using a cut of $\sigma >
150$ \kms, the value decreases by less than 4\%. However, we caution
against adopting such a correction because it does not appear to be
statistically significant. If we define the fundamental plane zero
points of the distant clusters by minimizing the absolute residuals,
the derived value of $H_0$ decreases by less than 1\%, consistent with
the Monte Carlo simulations of the previous paragraph.

This robustness of our result is also illustrated using the cluster
with the largest sample, Coma, and varying the cut in velocity
dispersion. When we use the mean zero point of the Coma fundamental
plane, with no velocity dispersion cut, we obtain $H_0=82\pm
6$\kms\Mpc-1 (random). Using only those galaxies with $\sigma > 150$
\kms, and $\sigma > 200$ \kms, one obtains $H_0=80 \pm 6$ \kms\Mpc-1,
and  $H_0=81 \pm 6$ \kms\Mpc-1, respectively. When we minimize the
absolute residuals in determining the fundamental plane zero point of
Coma, we find $H_0=81\pm 6$ \kms\Mpc-1, using only those galaxies with
either $\sigma >150$ \kms or $\sigma >200$ \kms. This test is
consistent with the Monte Carlo simulations, in which the magnitude of
the bias is smaller than 1\%.

Based on these tests we argue that the cluster population
incompleteness bias is not a statistically significant systematic
effect in our analysis of the J\o{}rgensen \etal\ (1996) data.

We also test the magnitude of the bias in the calibrators by imposing
similar cuts on velocity dispersion, in Leo I, Virgo, and Fornax. The
net effect on the inferred value for $H_0$ is $\pm 1\%$. Using the
$\sigma>150$ \kms cut, $H_0$ decreases by 1\%. For a cut of $\sigma >
200$ \kms,  $H_0$ increases by 1\%.

Thus, using the experiments listed above with all 11 clusters, with
Coma alone, and with the nearby sample, we estimate that the cluster
population incompleteness bias introduces an uncertainty of at most
$\pm 2\%$.


\subsection{The Spatial Coincidence of the Spirals and Ellipticals}
\label{gonz}

Up until now, we have assumed that the early-type galaxies are
co-spatial with spirals for which we have Cepheid distances. If this
assumption is correct, there may still be depth effects in the three
groups/clusters. For extended clusters like Virgo, there is an increased
probability that galaxy targets are systematically selected from the
near side of a given cluster, while the ellipticals are selected from
the cluster core. Gonzalez \& Faber (1997) modeled the effects of this
bias for the Virgo cluster, and argued that the Key Project distance to
Virgo may be under-estimated by 5-8\% when using the Cepheid distance to
M100 alone.

In this paper we now have 11 Cepheid distances to 3 clusters whose
distances are uncertain to 2--7\%, compared to the original $\pm$10\%
distance to M100. The situation has clearly improved, but the potential
for bias is still there. We have therefore carried out simulations of
the type done by Gonzalez \& Faber in order to quantify the bias. Now
that we have 11 Cepheid distances, we can characterize the distance
dispersion in the clusters, averaging over the three of them after
appropriate scaling. This is appropriate, as there is clearly
line-of-sight elongation in both the Virgo and Fornax spiral population,
(and insufficient information on Leo I). The distance dispersion works
out at 0.38 mag in distance modulus. The ellipsoidal model of Gonzalez
\& Faber, sampled with the standard conical field-of-view, predicts such
a dispersion with major axis scale length, $b$ = 3 Mpc, when the minor
axis scale length, $a$ = 1 Mpc. Gonzalez \& Faber assumed $a$ = 1, and
$b$ = 2.5--4 Mpc. Our 11 galaxy sample has allowed us to narrow the
range of $b$ in the cluster model. For fields-of-view of diameter
between 8$^\circ$ and 12$^\circ$, the bias due to the geometrical
extension of the cluster is 0.08 to 0.10 mag (4--5\% in distance), in
the sense that underestimated cluster distances will tend to bias H$_0$
high.

Gonzalez \& Faber also noted another potential source of bias, the
selection of the Virgo cluster itself. This selection results in a
Malmquist bias of the conventional type present in homogeneous
distributions of sampled objects (in this case galaxy clusters). Now
that we have 3 clusters with distance uncertainties 0.04--0.14 mag,
correction for this bias (by 1.38$\sigma^2$ mag) can be neglected. As
Gonzalez \& Faber foresaw, ``this particular bias correction is only
of temporary interest, as Virgo will soon cease to be the lynch-pin of
the Key Project.''

In summary, we adopt a 5\% downward correction, as determined from the
modeling described above. Given the uncertainties in the line-of-sight
structure of Virgo and Fornax, we assume the uncertainty in this
correction is also $\pm 5\%$. This uncertainty may be conservative,
given that our estimate for the bias is in agreement with values
determined empirically by Tonry \etal\ (1999) and Ferrarese \etal\
(1999).


\section{The Error Budget and the Final Value of the Hubble Constant}
\label{errors}

The goal of the Key Project is to measure a value for $H_0$ with a
realistic uncertainty at or below 10\%. In Table \ref{tab:errors}, we
list an extensive error budget for this determination of the Hubble
constant. The table has three sections, outlining the various stages
of the measurement of $H_0$ using the fundamental plane and \dns. In
the first stage, we outline the typical error budget for a Cepheid
distance measurement. The systematic errors in the Cepheid distance
scale propagate directly to the value of $H_0$. These systematics
include the uncertainty in the LMC distance, and the WFPC2 photometric
zero points, which have been used in nearly all of the Key Project
distance determinations. The total systematic error in the Cepheid
distance scale amounts to $\pm 7\%$ in $H_0$. The random errors in the
Cepheid distances are treated as uncorrelated from galaxy to galaxy,
and thus decrease by $\sqrt{N}$, when using $N$ galaxies to measure
the distance to a group or cluster. The individual random errors in
the Cepheid distances used here were listed in Table \ref{tab:calibs},
and are combined to provide the random errors in the distances to Leo
I, Virgo, and Fornax given in Table \ref{tab:errors}.

The second section of the error budget lists components from each step
in the derivation of the local fundamental plane and \dns\ zero
points. In the following sections, we attempt to quantify all the
sources of error which propagate into uncertainties in our derived
value of $H_0$. In particular, we estimate the impact on $H_0$ arising
from systematic errors in the velocity dispersions and photometry of
the calibrating galaxies, with respect to equivalent measurements made
of the distant cluster samples. We also discuss the uncertainties due
to the peculiar velocity field in the distant sample, and
uncertainties due to errors in the fundamental plane slope.

Following the detailed discussion, we summarize the total
uncertainties and provide the final, corrected value for the local
expansion rate.

\subsection{Velocity Dispersions}

The largest single source of systematic error in the fundamental plane
component of our analysis is in the comparison of the velocity
dispersions of the calibrating galaxies with the distant ones.
Fortunately, the exponent on $\sigma$ in the fundamental plane is
small, of order unity, and small systematic errors in $\sigma$ remain
as small systematic errors in $H_0$. This aspect can be contrasted to
the Faber-Jackson and Tully-Fisher relations (Faber \& Jackson 1976,
Tully \& Fisher 1977), for which the exponents are 3-4, causing errors
in the internal kinematics to lead to large errors in derived
distances.

J\o{}rgensen \etal\ (1995a) attempted to homogenize the wealth of
velocity dispersion information in the literature and made a detailed
study of the systematic differences between existing kinematic
datasets. They found that, in general, the Dressler \etal\ (1987)
measurements obtained at LCO were quite good, with small systematic
offsets between Dressler \etal\ and J\o{}rgensen \etal. The mean
offsets for the two sets of LCO dispersions were equivalent to $+3\%$
and $-4\%$ in distance, with \rms\ scatter of $\pm 7\%$ and $\pm 9\%$.
Such scatter leads to uncertainties in the fundamental plane
zero points of $\pm 2\%$ and $\pm 3\%$. The velocity dispersions of
the Virgo and Fornax galaxies used in this paper were derived
primarily from LCO observations so we anticipate that these
uncertainty estimates are appropriate for our analysis. The Fisher
(1997) measurements in Leo I were performed with the same
algorithm as the J\o{}rgensen \etal\ (1995a) system and thus are
expected to have negligible systematic uncertainties as well. We expect
that any resulting systematic errors due to the differences in the
systems of velocity dispersions are going to be at the level of 3-4\%.

The aperture corrections we applied were generally small, but small
uncertainties may remain and lead to errors in $H_0$. The Dressler
\etal\ (1987) dispersions were measured in $16''\times 16''$
apertures, equivalent to a circular aperture of $D=18\Sec 5$
(J\o{}rgensen \etal\ 1995a). For the Virgo galaxies, this aperture is
equivalent to $D=3\Sec 3$ at the distance of Coma; for the Fornax
galaxies it is $D=4\Sec 3$. The \rms\ scatter in the J\o{}rgensen
\etal\ (1995) modeling indicates that the uncertainties in the
aperture corrections for Virgo and Fornax are less than 1\%.
(J\o{}rgensen 1999, private communication). For the two Fisher (1997)
galaxies, we applied aperture corrections of about 9\%. The \rms\
scatter in the J\o{}rgensen \etal\ (1995) aperture corrections
indicates that the uncertainty in the aperture corrected velocity
dispersion for such large corrections is less than $\pm 6\%$ per
galaxy. Dividing by $\sqrt{2}$, we find a net effect of 4\%. This
uncertainty is diluted by the addition of the Faber \etal\ (1989)
galaxies, whose velocity dispersions were obtained through apertures
larger by a factor of three. Using these two galaxies, the net
uncertainty due to the aperture corrections is less than 3\%. The
total uncertainty in the Leo I Fundamental Plane zero point due to
aperture corrections is less than $\pm 0.02$ dex. Averaged over the
three groups/clusters, the net error due to the aperture corrections
is likely to be on the order of, or less than $\pm 0.01$ dex.

The level of rotational support in these galaxies can contribute to the
scatter in the fundamental plane, but will not lead to systematic errors
because the aperture corrections implicitly include both the gradient in
velocity dispersion, and the variation of rotational support with
radius. The gradient in the {\it total\/} second moment of the
line-of-sight velocity distribution is a regular, smooth function, with
little actual dependence on the level of rotational support and galaxy
morphology, unlike the gradients in $\sigma$ alone (Kelson \etal\
1999c). Furthermore, the ranges of rotational support seen in each
cluster are broadly consistent with each other ({\it e.g.\/}, Bender,
Saglia, \& Gerhard 1994, D'Onofrio \etal\ 1995, Fisher 1997). We
conclude that errors in $H_0$ due to uncertainties in the aperture
corrections are most likely small at a \about 2\%.


\subsection{Structural Parameters}

Differences in measurement techniques can lead to large random errors
in the individual measurements of effective radii and surface
brightnesses. The systematic offsets between these measurements made
by different authors are typically small, at the level of few percent,
despite potentially large uncertainties in the true half-light radius
of a given galaxy (greater than $50\%$ is not uncommon at times).
Nevertheless, the combination of $\theta_e$ and $\langle I\rangle_e$
that appears in the fundamental plane is very stable given that the
correlation coefficient between $\log \theta_e$ and $\log \langle
I\rangle_e$ is $-0.73$. The systematic errors in this quantity,
evidenced by differences between various authors, are typically at the
level of a few percent, depending on the quality of the
photometry/imaging data. The random errors are typically a few percent
as well, tested by fitting different profile shapes to galaxies and
re-deriving values of \fpparam{-0.82}.


\subsection{Photometry}


There are several sources of uncertainty which can arise in the
derivation of the Gunn $r$ surface brightnesses themselves. These
include the initial calibration by Tonry \etal\ (1997), uncertainties in
the transformation to Gunn $r$, errors in the galaxy colors, or in the
assumption that color gradients in the galaxies are negligible. Each of
these sources of error are discussed below in the context of our error
budget.


\subsubsection{Calibration}

The Tonry \etal\ (1997) photometric system referenced to the system of
Landolt (1992), the same system referred to by J\o{}rgensen (1994), who
explicitly derived transformations from the Landolt (1992)
Johnson-Kron-Cousins system to the Thuan \& Gunn (1976) photometric
system. This suggests that systematic errors between the photometric
systems of Tonry \etal\ (1997) and J\o{}rgensen (1994) are very small.
Tonry \etal\ (1997) quote uncertainties in the $V$ photometric zero
point of $\pm 2\%$ and errors in $(V-I_c)$ of $\pm 2\%$ and these are
likely to propagate into systematic uncertainties in the Hubble
constant.


\subsubsection{Transformation to Gunn $r$ Photometry}

J\o{}rgensen (1994) reported an $\rms$ scatter in their photometric
transformation of approximately $\pm 0.02$ mag. We can test the validity
of the transformation by reversing the direction of the transformation,
and re-measuring the Hubble constant. Earlier, the $V$ and $I_c$ colors
were used to transform the Leo I, Virgo, and Fornax data to Gunn $r$,
for direct comparison to the J\o{}rgensen \etal\ database. However, a
subset of 28 galaxies in Coma also have $B-r$ colors. Thus, we can
transform these to Johnson $V$, for direct comparison with the Leo I,
Virgo, and Fornax galaxies. In this way, we explicitly test the
systematic uncertainties of Eq. \ref{eq:xfm}, as well as our
approximation of $(B-V) \approx 0.814 (V-I_c)$. The net effect on the
Hubble constant is less than 2\%. Thus, we adopt an error estimate of
2\% for our error budget. We note that this test is not independent of
the accuracy of the $B-r$ colors of the J\o{}rgensen \etal\ Coma
galaxies.


\subsubsection{Consistency of Galaxy Colors}

A subset of 28 Coma galaxies have both Gunn $r$ and Johnson $B$
photometry. Using these data, we find that the mean $V-r$ colors of
Coma early-type galaxies is $\langle V-r\rangle= 0.198\pm 0.002$ mag
(internal error only). We can compute the mean $V-r$ colors of the Leo
I, Virgo, and Fornax galaxies using Eq. \ref{eq:xfm} and rewriting it
as
\begin{equation}
(V-r) = -0.273 + 0.396\times (V-I_c)
\label{eq:xfm2}
\end{equation}
We report mean $(V-r)$ colors for the galaxies in Leo I, Virgo, and
Fornax of $0.177\pm 0.013$ mag, $0.185\pm 0.005$ mag, $0.195\pm 0.005$
mag (again, these are the internal errors). The uncertainties on the
colors themselves are on the order of $\pm 0.01$ to $\pm 0.02$ mag. This
similarity in the galaxy colors suggest a remarkable uniformity in the
stellar populations ({\it e.g.\/}, Bower, Lucey, \& Ellis 1992). Given
the consistent galaxy colors, we conclude that the Gunn $r$ surface
brightnesses are likely to be good to $\pm 0.02$ mag.


\subsubsection{The Effect of Color Gradients}
\label{errgrad}

One issue that has not been addressed is the assumption that color
gradients are negligible. In \S \ref{findrn}, we noted that the color
gradients were indeed small. We can explicitly test their effects by
measuring the structural parameters in the $I_c$ band, transforming
the $I_c$-band surface brightnesses to Gunn $r$ using the $(V-I_c)$
colors, and re-deriving the fundamental plane zero points. For the
Virgo galaxies, there is less than a 1\% change in the zero point of
the fundamental plane. For the Fornax galaxies, there is a net change
of 2.5\%. This test is not independent of the any errors in the
photometric transformation, but given the previous discussions, none
of these uncertainties lead to large errors in $H_0$.


\subsection{Systematic Uncertainties due to the Velocity Field}
\label{pecvel}

Using the fundamental plane, we derived identical values for the Hubble
constant using recession velocities for the distant clusters in both the
CMB reference frame, and in the frame of the flow-field model of Mould
\etal\ (1999). In order to establish our estimate of the error due to
uncertainties in the velocity field, we studied those distant clusters
with more than 20 galaxies and found a standard deviation of 4\% in
their implied values of the Hubble constant. This scatter is a
combination of the random uncertainties in the distant zero points due
to the finite numbers of galaxies, and the errors due to the random
motions of the clusters on top of the smooth Hubble flow. We
conservatively adopt the 4\% as the random error in a measurement of the
Hubble constant from a single distant cluster. We assume that this
uncertainty decreases with the square-root of the number of distant
clusters and adopt an error in final determination of $H_0$ of 1\%.


\subsection{Uncertainties in the Slope of the Fundamental Plane and \dns\
Relations}
\label{sloperr}

The fundamental plane is essentially a relation between $M/L$ ratio
and galaxy mass (Faber \etal\ 1987). In order for the fundamental
plane to be effective as a distance indicator, this scaling between
$M/L$ and the structural parameters must be consistent from cluster to
cluster. If $M/L$ scales differently in any of the clusters, then for
a given value of $\sigma$ and $\theta_e$, the predicted $M/L$ will not
be correct, and the deduced distance will be incorrect as well. We now
consider uncertainties in the fundamental plane slope, and their
impact on our value of the Hubble constant.

Varying the slope on $\log\sigma$ from 1.14-1.30 (the $\pm 1$-$\sigma$
range in J\o{}rgensen \etal\ 1996) amounts to a 1\% variation in the
Hubble constant. The published uncertainty in $\beta$ of $\pm 0.02$
also leads to negligible changes in the Hubble constant.

However, there remains an ambiguity in that we have two scaling
relations, the FP and \dns, which are not equivalent in shape but have
similar scatter ({\it e.g.\/}, $\pm 10\%$ in distance in Virgo). We
also noted that there is a specific relationship
$\theta_n=f(\theta_e,\langle I\rangle_e)$ with a slope which is
different from the value of $\beta$ in the fundamental plane. Thus, we
now experiment with revised forms of the fundamental plane by fixing
$\beta=-0.73$, the value implicit in \dns.

Taking the Coma data of J\o{}rgensen \etal\ (1995ab) and performing a
linear least squares fit, to those galaxies with $\log \sigma >2$, we
find the following \dns\ relation
\begin{equation}
\log \theta_n \propto (1.22\pm 0.08)\log\sigma
\end{equation}
This relation corresponds to a fundamental plane of the form $\log
\theta_e = 1.22\log\sigma-0.73\log\langle I\rangle_e$.

Using this new plane, we find (metric) zero points for Leo I, Virgo,
and Fornax of $\langle \gamma\rangle_{\rm Leo I}=-0.309\pm 0.042$,
$\langle \gamma\rangle_{\rm Virgo}=-0.371\pm 0.013$, $\langle
\gamma\rangle_{\rm Fornax}=-0.345\pm 0.033$. The new (angular) zero
point for Coma is $\langle\gamma\rangle_{\rm Coma}=-5.325\pm 0.010$.
The net effect on the Hubble constant is $-2.5\%$.

The small impact on the Hubble constant that arises when the
fundamental plane coefficients are adjusted is largely due to the
averaging of the zero points in the three clusters. By setting the
zero point to only one of the clusters alone, values of $H_0$ may vary
over a slightly broader range. The Fornax cluster, with its large
scatter, is the worst case. Given the large scatter in Fornax, a
modest change in $\beta$ will lead to larger zero point shifts because
of the correlation between $\theta_e$ and $\langle I\rangle_e$. The
shift from $\beta=-0.82$ to $\beta=-0.73$ decreases its Hubble
constant by 6\%. Given the scatter in the Fornax FP, this offset is
not statistically significant.

Averaging over the Leo I, Virgo, and Fornax diminishes the uncertainty
in $H_0$ due to errors in the FP slopes primarily because the scatter in
the Virgo relation is so low. Nevertheless, we conclude that
uncertainties in the fundamental plane slopes are at most a small
contributor to the total error in the Hubble constant itself.


\subsection{The Total Error in $H_0$}
\label{totalerr}

Given the level of uncertainties listed in Table \ref{tab:errors}, an
accuracy of $\pm 10\%$ cannot be achieved using the fundamental plane
or \dns\ relations alone. The largest single sources of error are
currently systematic: (1) the LMC distance, (2) the WFPC2 photometric
calibration, (3) the Cepheid metallicity correction, and (4) depth
effects between the spirals and ellipticals. The systematic errors in
the Cepheid distance scale, including the uncertainty in the LMC
distance, combine to a total of $\pm 7\%$. The random errors in the
Cepheid distances do not contribute much to the total random error in
$H_0$; it is the superposition of the random small errors in the FP
(and \dns) which make up the bulk of the total random uncertainty of
$\pm 6\%$. Remaining systematic uncertainties arising from the
fundamental plane analysis and the systematic corrections, when
combined in quadrature, amount to about $\pm 8\%$, leading to a total
systematic uncertainty in $H_0$ of $\pm 11\%$.

However, by combining several independent distance indicators, the
goal of the Key Project may be realized (Mould \etal\ 1999). Each
method provides its own set of constraints and systematic
uncertainties on the value, thereby reducing the total uncertainty in
the Hubble constant. This subject is the focus of Mould \etal\ (1999a)
and Freedman \etal\ (1999), who attempt to combine all of the
constraints provided by this work, Ferrarese \etal\ (1999), Gibson
\etal\ (1999), and Sakai \etal\ (1999).

\subsection{A Final Value for the Hubble Constant}
\label{final}

In \S \ref{fplocal}, we used the fundamental plane in Leo I, Virgo,
and Fornax to set the distance scale. Upon comparing the zero points
of the distant clusters to the one defined by the calibrating
galaxies, we determined a raw value of $H_0 =82\pm 5 \pm 10$
\kms\Mpc-1. However, we also estimated to what extent the value of
$H_0$ is likely to be biased by the assumption that the Key Project
spirals may be used to set distances to the three clusters used in
this paper, and concluded that the raw value for the expansion rate is
likely to be biased by approximately $-5\%\pm 5\%$. We therefore
correct the estimate of $H_0$ derived in \S \ref{fplocal} and arrive
at our adopted value of $H_0 = 78 \pm 5 \pm 9$ \kms\Mpc-1. Adoption of
the Kennicutt \etal\ (1998) metallicity correction, would further
reduce this value to $H_0=73\pm 4\pm 9$ \kms\Mpc-1.


\section{Comparisons with the Literature}
\label{complit}

Our value of $H_0$ using Leo I alone can be compared directly with
that of Hjorth \& Tanvir (1997), who used a Cepheid distance to NGC
3368 of $11.6\pm 0.9$ Mpc. Our distance is 5\% shorter. Had Hjorth \&
Tanvir (1997) adopted the Key Project distance to Leo I, and the
flow-field velocity of $cz_{flow}=7392$ \kms, they would have found
$H_0=77\pm 9$ \kms\Mpc-1 instead of $H_0=67\pm 8$ \kms\Mpc-1. Using
Leo I alone and Coma, we find $H_0=70\pm 4$ \kms\Mpc-1 (internal).
This 10\% discrepancy can arise from either the photometry or the
velocity dispersions. Comparing values of $\theta_e\langle
I\rangle^{0.82}$ in Gunn $r$, our values differ from those of Hjorth
\& Tanvir (1997) by only 1\% in the mean. This comparison is
remarkable considering our error estimate of 5\% in Table
\ref{tab:errors}.

The discrepancy between the Hjorth \& Tanvir (1997) result and ours can
be attributed solely to a systematic difference in the
aperture-corrected velocity dispersions. For NGC 3377 and NGC 3379, the
comparison is quite good; our aperture-corrected velocity dispersions
agree to 3\%. For NGC 3384 and NGC 3412, the comparison is quite poor;
our aperture-corrected values are 20\% lower than Hjorth \& Tanvir's
estimates. According to our error budget, we estimated that the total
systematic uncertainty in the velocity dispersions was of order 5\%.
Perhaps the correction for apertures as small as those used by Fisher
(1997) is uncertain by more than we estimated. J\o{}rgensen \etal\
(1995) approximated the correction from small-aperture velocity
dispersions to equivalent large-aperture measurements using a sample of
51 nearby galaxies. The power-law they derived is the median relation
for that sample, and we used the \rms\ scatter about this relation to
estimate our uncertainties in \S \ref{errors}. Perhaps the uncertainties
in the Hjorth \& Tanvir (1997) estimated corrections are larger than
those authors assumed.

We can also compare our results to that of Gregg (1995). Using the
$K$-band \dns\ relation, he inferred a relative Coma-Virgo distance of
5.56 (no error estimate). Using the fundamental plane, we find
$5.57\pm 0.50$, and using \dns, we find $5.87\pm 0.52$, both of which
are in excellent agreement.

Another test of our analysis can be made by comparing our structural
parameters with those of Faber \etal\ (1989), who reported values of
$A_e$ (effective diameter $\equiv 2\times \theta_e$) and $\Sigma_e$
(the $B$-band surface brightness within $A_e$) for a large sample of
early-type galaxies, include several in Leo I, Virgo, and Fornax.
Using $(B-V)\approx 0.814\times (V-I_c)$ and the J\o{}rgensen (1994)
transformation,  we can correct the $B$-band surface brightnesses to
Gunn $r$, and compare fundamental planes.

Faber \etal\ only report data for two galaxies in Leo I, N3377 and
N3379. Using these two galaxies, there is a systematic difference
between our photometry and theirs of $-4\%$ in distance. However, given
only two galaxies, it is difficult to evaluate the significance of this
discrepancy. For the Virgo galaxies, the comparison is quite favorable.
The Faber \etal\ (1989) data, transformed to Gunn $r$, produce a
fundamental plane zero point only offset $-3\%$ in distance from the one
derived earlier from the Tonry \etal\ (1997) photometry. This offset is
statistically insignificant. For the Fornax galaxies, the offset is
$-11\%\pm 4\%$ (internal error only). In \S \ref{errors}, we estimated
that our fundamental plane parameters may have uncertainties on the
order of $\pm 5\%$. We have no estimates for the uncertainties in the
Faber \etal\ (1989) photometry, so the errors on the Fornax offset given
above is a lower limit. We also note that the scatter in the Virgo and
Fornax fundamental planes, derived from the Faber \etal\ data is
identical to what was derived earlier using the Tonry \etal\ (1997)
photometry.

D'Onofrio \etal\ (1997) used the $B$ band \dns\ and fundamental plane
relations to determine the Virgo-Fornax relative distance modulus.
Using the fundamental plane, they found a relative modulus of
$\mu_{FV}=0.45\pm 0.15$ mag. The fundamental plane zero points we find
for Virgo and Fornax imply $\mu_{FV}=0.52\pm 0.17$ mag. Given this
agreement, we conclude that the disagreement between our data and
Faber \etal\ (1989) for the Fornax galaxies may point to larger
uncertainties in the Faber \etal\ (1989) dataset than anticipated. We
also note that D'Onofrio \etal\ (1997) find values for the scatter in
the Virgo and Fornax fundamental plane relations that are similar to
ours.


\section{Conclusions}

Using the fundamental plane relations in Leo I, Virgo, and Fornax and
comparing the 11 clusters of J\o{}rgensen \etal\ (1995), we found a
raw value for the local Hubble constant of $82\pm 5\pm 10$ \kms\Mpc-1
(random, systematic). Correcting for depth effects in the nearby
clusters, we estimate that the local expansion rate is $5\% \pm 5\%$
smaller and conclude that $H_0=78 \pm 5\pm 9$ \kms\Mpc-1.

The largest systematic uncertainties come from the LMC distance, the
WFPC2 calibration, the metallicity correction to the two-color Cepheid
distances, and the uncertainty in the Cepheid distances to the three
nearby clusters. By adopting the Kennicutt \etal\ (1998) metallicity
correction the distances to Leo I, Virgo, and Fornax decrease by $8\%\pm
5\%$, $6\%\pm 4\%$, and $4\%\pm 3\%$, respectively. The net effect would
be to decrease the Hubble constant by $6\%\pm 4\%$ to $H_0=73 \pm 4 \pm
9$ \kms\Mpc-1.

The distant cluster sample spans a wide range of recession velocities,
from $cz\approx 1100$ \kms to $cz\approx 11000$ \kms, with a mean $cz
\approx 6000$ \kms. The value of $H_0$ using either the CMB, or the
flow-field model of Mould \etal\ (1999a) as the preferred reference
frame, give the same value to within 1\%. Using Coma with its CMB
velocity, yields a value for the local expansion rate which is nearly
4\% smaller. However, using the flow-field reference frame, Coma
yields the same value of the Hubble constant as the weighted mean of
the 11 clusters, to within less than 1\%. Therefore, by using all 11
clusters of J\o{}rgensen \etal\ (1996), we anticipate that
uncertainties due to the peculiar velocity field are likely to be at a
level of 1\%.

When using the \dns\ relation, we find a value which is smaller by
4\%, but the \dns\ relation is a curved projection of the fundamental
plane (J\o{}rgensen \etal\ 1993, and this paper). Thus, the
fundamental plane is preferred and appears to be a more accurate
secondary distance indicator, particularly when care is taken to
minimize, and quantify the systematic errors.

The values of $H_0$ given above were derived by fixing the fundamental
plane and \dns\ relations' zero points to the Cepheid distances to Leo
I, Virgo, and Fornax spirals. The Cepheid distance to Fornax is
defined by NGC 1365, NGC 1326A and NGC 1425. The distance modulus to
NGC 1425 is 0.4 mag larger than the former two. Suntzeff \etal\ (1999)
have suggested that the hosts of Ia supernovae may lie several tenths
of a magnitude beyond NGC 1365. Thus, if the ellipticals used in this
paper are at the distance of NGC 1425, then the Hubble constant,
derived using a weighted mean of the three calibrating clusters,
decreases by 2\%. If the Fornax ellipticals are at the distance of the
nearer of the two Cepheid distances then the Hubble constant increases
by about 2\%.

While our uncertainty on the Hubble constant ($H_0=78 \pm5\pm
9$\kms\Mpc-1) from this single distance indicator is greater than 10\%,
these results will be combined by Mould \etal\ (1999a) and Freedman
\etal\ (1999) with results from the Type Ia Supernovae (Gibson \etal\
1999), Tully-Fisher relation (Sakai \etal\ 1999), and surface-brightness
fluctuation method (Ferrarese \etal\ 1999) to produce a more accurate
value of the local expansion rate of the Universe.


\acknowledgements

We gratefully the advice of A. Gonzalez in estimating the uncertainty in
the spatial coincidence of spirals and ellipticals. The work presented
in this paper is based on observations with the NASA/ESA Hubble Space
Telescope, obtained by the Space Telescope Science Institute, which is
operated by AURA, Inc. under NASA contract No. 5-26555. We gratefully
acknowledge the support of the NASA and STScI support staff, with
special thanks to our program coordinator, Doug Van Orsow. Support for
this work was provided by NASA through grant GO-2227-87A from STScI.
SMGH and PBS are grateful to NATO for travel support via Collaborative
Research Grant 960178. LF acknowledges support by NASA through Hubble
Fellowship grant HF-01081.01-96A awarded by the Space Telescope Science
Institute, which is operated by the Association of Universities for
Research in Astronomy, Inc., for NASA under contract NAS 5-26555. This
research has made use of the NASA/IPAC Extragalactic Database (NED)
which is operated by the Jet Propulsion Laboratory, California Institute
of Technology, under contract with the National Aeronautics and Space
Administration.



\clearpage

\clearpage

\begin{center}
\begin{minipage}{\textwidth}
\vspace{2.3in}
\centerline{\mkfigbox{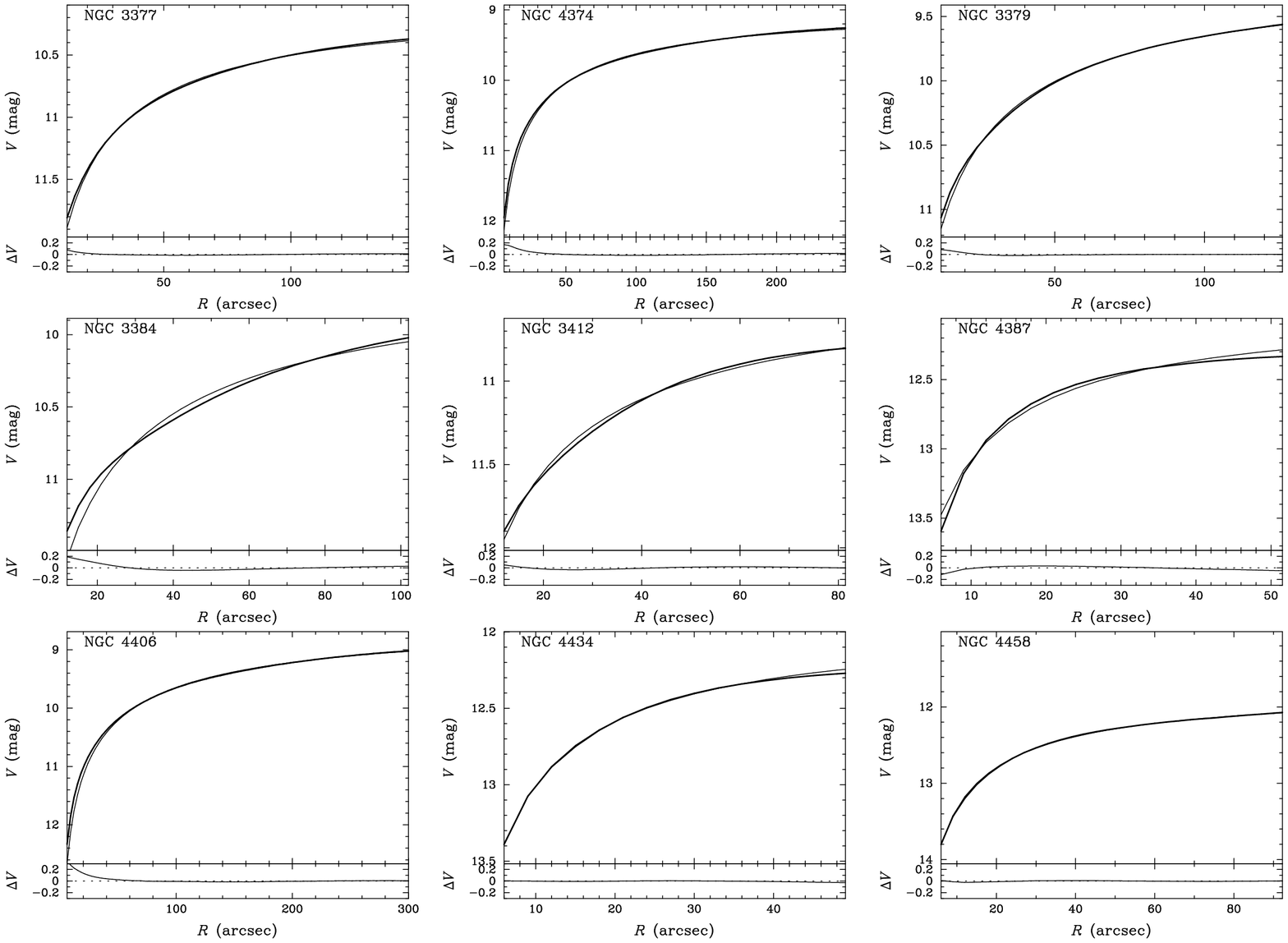}{0.8\textwidth}}
\vspace{0.15in}
\Fig
\end{minipage}
\end{center}
\clearpage

\begin{center}
\begin{minipage}{\textwidth}
\vspace{2.3in}
\centerline{\mkfigbox{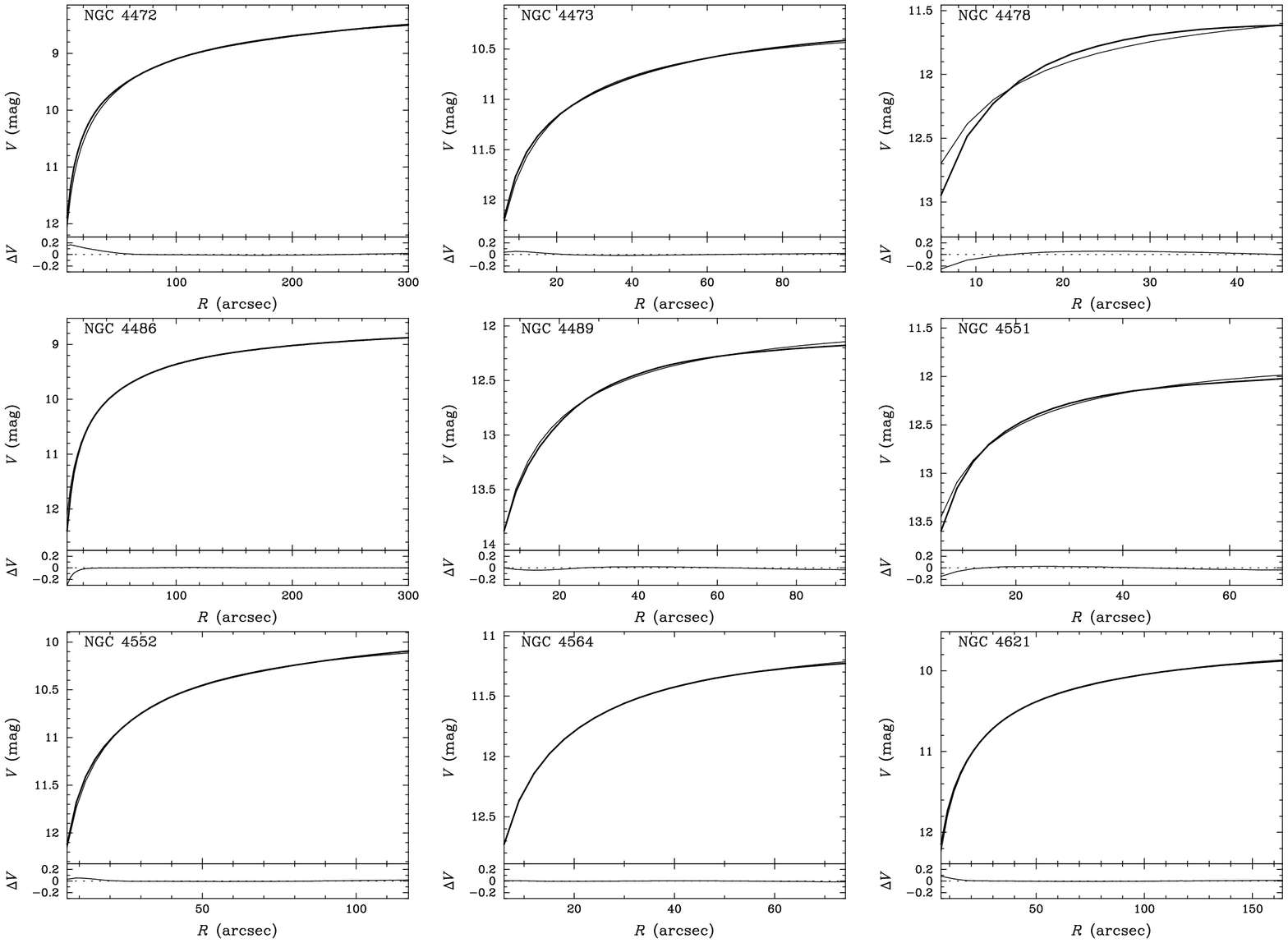}{0.8\textwidth}}
\vspace{0.15in}
\Fig
\end{minipage}
\end{center}
\clearpage

\begin{center}
\begin{minipage}{\textwidth}
\vspace{2.3in}
\centerline{\mkfigbox{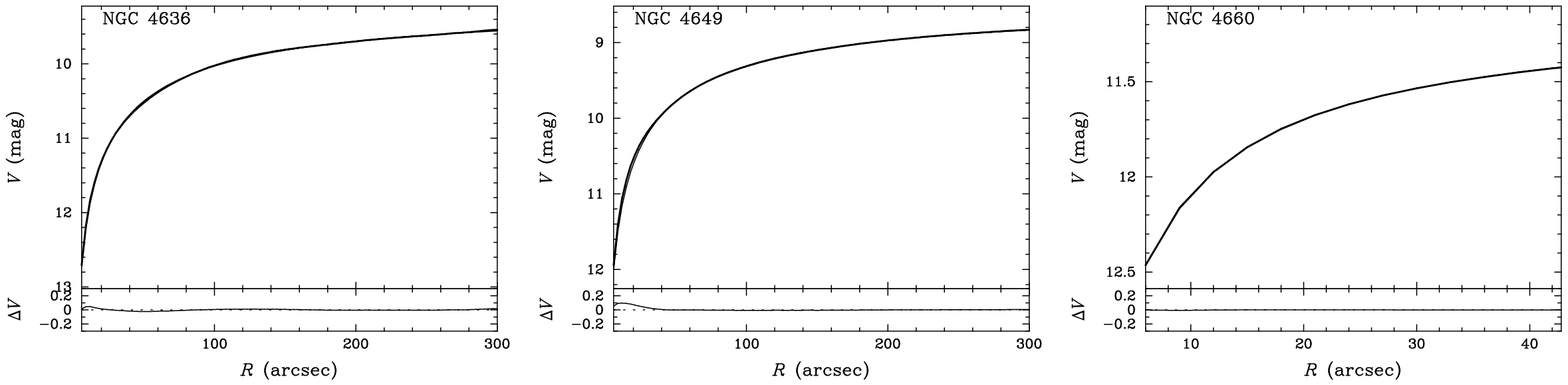}{0.8\textwidth}}
\vspace{0.15in}
\Fig
\end{minipage}
\end{center}
\clearpage

\clearpage

\clearpage


\begin{deluxetable}{l l r c c c c c l r}
\tiny
\tablewidth{0pt}
\tablecaption{Fundamental Plane and \dns\ Quantities \label{tab:params}}
\tablehead{
\colhead{} &
\colhead{} &
\multicolumn{2}{c}{$V$-band} &
\colhead{$r$-band} &
\colhead{} &
\colhead{} &
\colhead{} \nl
\colhead{Galaxy} &
\colhead{Type} &
\colhead{$\theta_e$$^a$} &
\colhead{$\langle \mu\rangle_e$$^a$} &
\colhead{$\theta_n$} &
\colhead{$\log \langle I\rangle_e$$^b$} &
\colhead{$\log \sigma$$^c$} &
\colhead{$(V-I_c)$} &
\colhead{$E(B-V)$} &
\colhead{$cz$$^d$} \nl
\colhead{} &
\colhead{} &
\colhead{(arcsec)} &
\colhead{(mag/arcsec$^2$)} &
\colhead{(arcsec)} &
\colhead{$L_\odot/pc^2$$^e$} &
\colhead{(km/s)} &
\colhead{(mag)} &
\colhead{(mag)} &
\colhead{(km/s)}
}
\startdata
\nl
NGC 3377     &E5     & $ 41.78\pm 0.53$ & $20.28\pm 0.02$ & 32.19 & 2.572 & 2.100 & 1.22 & 0.03& 692 \nl
NGC 3379     &E1     & $ 50.40\pm 0.64$ & $19.76\pm 0.03$ & 54.21 & 2.776 & 2.287 & 1.21 & 0.02& 920 \nl
NGC 3384     &SB0    & $ 49.91\pm 1.16$ & $20.18\pm 0.05$ & 42.20 & 2.609 & 2.101 & 1.17 & 0.03& 735 \nl
NGC 3412     &SB0    & $ 23.27\pm 0.61$ & $19.54\pm 0.05$ & 28.55 & 2.857 & 1.974 & 1.18 & 0.03& 865 \nl
\nl                                                                       
NGC 4365$^f$ &E3     & $ 51.10\pm 0.84$ & $20.31\pm 0.03$ & 38.86 & 2.552 & 2.409 & 1.22 & 0.02 &  1240 \nl
NGC 4374     &E1     & $ 71.15\pm 0.53$ & $20.33\pm 0.02$ & 55.19 & 2.561 & 2.477 & 1.21 & 0.04 &  1000 \nl
NGC 4387     &E5     & $ 14.72\pm 3.65$ & $19.88\pm 0.49$ & 14.85 & 2.726 & 2.056 & 1.17 & 0.03 &  561  \nl
NGC 4406     &S0/E3  & $150.82\pm 0.80$ & $21.52\pm 0.01$ & 48.54 & 2.065 & 2.352 & 1.18 & 0.03 & -227  \nl
NGC 4434     &E0/S0  & $ 14.02\pm 3.76$ & $19.76\pm 0.53$ & 14.89 & 2.759 & 2.006 & 1.15 & 0.02 &  1071 \nl
NGC 4458     &E0     & $ 26.36\pm 4.22$ & $20.96\pm 0.32$ & 12.24 & 2.280 & 1.946 & 1.15 & 0.02 &  668  \nl
NGC 4472     &E2/S0  & $135.85\pm 0.43$ & $20.81\pm 0.01$ & 73.63 & 2.353 & 2.471 & 1.25 & 0.02 &  868  \nl
NGC 4473     &E5     & $ 27.64\pm 0.96$ & $19.43\pm 0.07$ & 36.69 & 2.905 & 2.265 & 1.18 & 0.03 &  2240 \nl
NGC 4478     &E2     & $ 12.92\pm 1.27$ & $18.96\pm 0.18$ & 23.38 & 3.085 & 2.167 & 1.17 & 0.03 &  1381 \nl
NGC 4486     &Epec   & $ 98.27\pm 0.41$ & $20.58\pm 0.01$ & 62.29 & 2.449 & 2.525 & 1.25 & 0.02 &  1282 \nl
NGC 4489$^g$ &E      & $ 26.40\pm 4.50$ & $21.03\pm 0.34$ & 11.34 & 2.245 & 1.775 & 1.08 & 0.03 &  971  \nl
NGC 4551     &E      & $ 19.96\pm 3.48$ & $20.24\pm 0.34$ & 16.02 & 2.589 & 2.018 & 1.18 & 0.04 &  1470 \nl
NGC 4552$^h$ &E      & $ 35.97\pm 0.81$ & $19.66\pm 0.04$ & 42.60 & 2.828 & 2.388 & 1.21 & 0.04 &  321  \nl
NGC 4564     &E6     & $ 21.17\pm 1.70$ & $19.63\pm 0.16$ & 25.31 & 2.833 & 2.182 & 1.20 & 0.04 &  1111 \nl
NGC 4621     &E5     & $ 46.82\pm 0.76$ & $20.02\pm 0.03$ & 43.57 & 2.675 & 2.335 & 1.20 & 0.03 &  424  \nl
NGC 4636     &E/S0   & $ 92.94\pm 0.76$ & $21.15\pm 0.02$ & 39.13 & 2.227 & 2.300 & 1.27 & 0.03 &  1095 \nl
NGC 4649     &E2     & $ 93.15\pm 0.40$ & $20.44\pm 0.01$ & 65.93 & 2.508 & 2.511 & 1.26 & 0.03 &  1413 \nl
NGC 4660$^f$ &E      & $ 12.25\pm 1.55$ & $18.64\pm 0.25$ & 26.02 & 3.225 & 2.259 & 1.20 & 0.03 &  1097 \nl
\nl                                                                       
NGC 1339     &E4     & $ 22.05\pm 3.04$ & $20.31\pm 0.27$ & 16.06 & 2.533 & 2.180 & 1.15 & 0.01 & 1367 \nl
NGC 1344     &E5     & $ 40.06\pm 1.21$ & $20.17\pm 0.06$ & 32.66 & 2.594 & 2.194 & 1.15 & 0.02 & 1169 \nl
NGC 1374     &E0     & $ 34.96\pm 2.19$ & $20.62\pm 0.12$ & 20.93 & 2.414 & 2.216 & 1.16 & 0.01 & 1352 \nl
NGC 1379     &E0     & $ 70.38\pm 3.97$ & $21.72\pm 0.12$ & 17.95 & 1.969 & 2.077 & 1.16 & 0.01 & 1380 \nl
NGC 1399     &E0     & $ 55.42\pm 0.94$ & $20.10\pm 0.04$ & 48.07 & 2.632 & 2.479 & 1.25 & 0.01 & 1447 \nl
NGC 1404     &E2     & $ 47.34\pm 1.08$ & $20.34\pm 0.05$ & 34.83 & 2.535 & 2.388 & 1.24 & 0.01 & 1942 \nl
NGC 1427     &E5     & $ 44.28\pm 2.25$ & $21.07\pm 0.10$ & 17.97 & 2.229 & 2.203 & 1.17 & 0.01 & 1416 \nl
IC 2006      &E      & $ 30.10\pm 2.84$ & $20.71\pm 0.19$ & 15.71 & 2.377 & 2.083 & 1.20 & 0.01 & 1364 \nl
\enddata
\tablecomments{
Hubble types taken from the RC3 and Ferguson (1989). The $V$-band
photometry listed here has not been corrected for galactic extinction,
nor for cosmological surface brightness dimming, nor have
$K$-corrections been applied. The FP and \dns\ zero points in Table
\ref{tab:mzpts} ($\gamma$, $\delta$) were computed after taking these
other effects into account. \nl
$^a$Errors in $\theta_e$ and $\langle \mu\rangle_e$ are strongly correlated,
such that their combination in the fundamental plane remains conserved
to a high degree of precision, even when errors in effective radius
and surface brightness are large.
\nl
$^b$Values of surface brightness are in units of solar luminosities per
square parsec in the Gunn $r$ band in the restframe of the galaxy.
\nl
$^c$Values for $\log \sigma$ have been corrected to a nominal
aperture of $D=3\Sec 4$ at the distance of Coma. Values of $\log
\sigma$ for NGC 3377 and NGC 3379 were taken from Faber \etal\ (1989), and
corrected using the more recent aperture corrections of J\o{}rgensen
\etal\ (1995a). The Virgo and Fornax $\log \sigma$ values are from
Dressler \etal\ (1987) who derived dispersions from $16'' \times 16''$
apertures. The values for NGC 3412 and NGC 3384 were taken from Fisher \etal\
(1996), and corrected for aperture size from their aperture of $2''
\times 4''$ to the nominal aperture.
\nl
$^d$Recession velocities, taken from NED, are used to compute the
$K$-corrections and cosmological surface brightness dimming.
\nl
$^e$$\langle I\rangle_e$, in solar luminosities per square parsec in
the Gunn $r$ band. Galactic exctinction has been removed.
$K$-corrections were performed using the individual redshifts. Surface
brightness dimming of $(1+z)^4$ has been removed using mean recession
velocities for Leo I, Virgo, and Fornax, of 750 \kms, 1050 \kms, and
1400 \kms, respectively.
\nl
$^f$SBF measurements of NGC 4365 and NGC 4660 indicate that they are not
associated with the other members of Virgo (Tonry \etal\ 1997). These
are disregarded in the FP and \dns\ analysis.\nl
$^g$Too blue; dropped from subsequent analysis.\nl
$^h$The image of the core of NGC 4552 was saturated so the observed
mean surface brightness and $\theta_n$ are in error; also dropped from
subsequent analysis.\nl
}
\end{deluxetable}


\begin{deluxetable}{l c c c l}
\tiny
\tablewidth{0pt}
\tablecaption{Cepheid Distance Moduli to Leo I, Virgo, and Fornax
Galaxies
\label{tab:calibs}}
\tablehead{
\colhead{Galaxy} &
\colhead{$(m-M)_0$$^a$ (mag)} &
\colhead{$(m-M)_0$$^b$ (mag)} &
\colhead{Random Error (mag)} &
\colhead{Original Reference$^c$}
}
\startdata
\cutinhead{Leo I}
NGC 3351   & $30.01$& 30.19 & $\pm 0.08$ &  Graham \etal\ (1997)\nl
NGC 3368   & $30.20$& 30.37 & $\pm 0.10$ &  Gibson \etal\ (1999)\nl
\nl
{\it weighted mean\/} & $30.08$& 30.27 & $\pm 0.11$ \nl
\cutinhead{Virgo}                       
NGC 4321   & $31.04$& 31.19 & $\pm 0.09$ &  Ferrarese \etal\ (1996)\nl
NGC 4496A  & $31.02$& 31.13 & $\pm 0.07$ &  Gibson \etal\ (1999)\nl
NGC 4535   & $31.10$& 31.27 & $\pm 0.07$ &  Macri \etal\ (1999) \nl
NGC 4536   & $30.95$& 31.03 & $\pm 0.07$ &  Gibson \etal\ (1999)\nl
NGC 4548   & $31.04$& 31.24 & $\pm 0.08$ &  Graham \etal\ (1999)\nl
NGC 4639$^d$& $31.80$& 31.92 & $\pm 0.09$ &  Gibson \etal\ (1999)\nl
\nl
{\it weighted mean\/} & $31.03$& 31.17 & $\pm 0.04$ \nl
\cutinhead{Fornax}                     
NGC 1326A  & $31.43$& 31.43 & $\pm 0.07$ &  Prosser \etal\ (1999)\nl
NGC 1365   & $31.39$& 31.50 & $\pm 0.10$ &  Silbermann \etal\ (1999)\nl
NGC 1425   & $31.81$& 31.93 & $\pm 0.06$ &  Mould \etal\ (1999b)\nl
\nl
{\it weighted mean\/} & $31.60$& 31.68 & $\pm 0.14$ \nl
\enddata
\tablecomments {
$^a$Distance moduli reported by the HST Key Project on the Extragalactic
Distance Scale. Random errors listed only.
$^b$Distance modulus corrected for metallicity using
$\Delta \mu_{VI_c}=-0.24$ mag/dex (Kennicutt
\etal\ 1998).
$^c$A full table of all Cepheid distance measurements is given in
Ferrarese \etal\ (1999).
$^d$NGC 4639 was excluded from the determination of the mean distance to
Virgo.
}
\end{deluxetable}
\clearpage


\begin{deluxetable}{l l l l l l}
\tiny
\tablewidth{0pt}
\tablecaption{Error Budget for Fundamental Plane and \dns
\label{tab:errors}}
\tablehead{
\colhead{} &
\colhead{} &
\colhead{Source} &
\colhead{} &
\colhead{Error$^a$} &
\colhead{Notes}
}
\startdata
\multicolumn{5}{l}{\bf 1. Errors in the Cepheid Distance Scale}\nl
&    & {\it A. LMC Distance Modulus\/}    && $\pm 0.13$ mag&
Adopted from Madore \& Freedman (1999)\nl
&    & {\it B. LMC P-L zero point$^b$\/}          && $\pm 0.02$ mag\nl
&S1.1& LMC P-L Systematic Error   && $\pm 0.13$ mag &
$\sqrt{A^2+B^2}$\nl
\nl
&    & {\it C. HST $V$-band zero point$^c$\/}    && $\pm 0.03$ mag\nl
&    & {\it D. HST $I_c$-band zero point$^c$\/}    && $\pm 0.03$ mag\nl
&S1.2& Systematic Error in the Photometry   && $\pm 0.09$ mag &
$\sqrt{C^2(1-R)^2+D^2 R^2}$, $R\equiv A_V/E(V-I_c)=2.47$ \nl
\nl
&R1.1& Random Error in the HST Photometry      && $\pm 0.05$ mag &
From DoPHOT/ALLFRAME comparisons\nl
\nl
&    & {\it E. Differences in $R$ between galaxy and LMC\/}
&& $\pm 0.014$ mag & See Ferrarese \etal\ (1998) for details \nl
&    & {\it F. Errors in the Adopted $R$ of LMC\/}
&& $\pm 0.02$ mag & See Ferrarese \etal\ (1998) for details \nl
&R1.2& Random Error in the Extinction Treatment      && $\pm 0.02$ mag &
$\sqrt{E^2+F^2}$\nl
\nl
&    & {\it G. Typical P-L zero point Error in $V$\/}          && $\pm 0.05$ mag\nl
&    & {\it H. Typical P-L zero point Error in $I_c$\/}          && $\pm 0.04$ mag\nl
&R1.3& Typical Random Error in the Cepheid P-L zero point&&
$\pm 0.06$ mag\nl
\nl
&R1$_T$& Typical Random Error for an Individual Cepheid Distance$^d$  && $\pm 0.08$ mag &
$\sqrt{R1.1^2 + R1.2^ 2+ R1.3^2}$\nl
\nl
&R1$_L$& Random Error in distance to Leo I && $\pm 0.11$ mag\nl
&R1$_V$& Random Error in distance to Virgo && $\pm 0.04$ mag\nl
&R1$_F$& Random Error in distance to Fornax && $\pm 0.14$ mag\nl
&S1& Total Systematic Error in the Cepheid Distance Scale&& $\pm 0.16$ mag &
$\sqrt{S1.1^2 + S1.2^ 2}$\nl
\nl

\multicolumn{5}{l}{\bf 2. Errors in the Fundamental Plane and \dns\
Relations}
\nl
&    & {\it I. Observed Velocity Dispersions\/}  && $\pm 0.01$ dex\nl
&    & {\it J. Aperture Corrections\/}  && $\pm 0.01$ dex\nl
&S2.1& Systematic Error from Velocity Dispersions&& $\pm 0.02$ dex &
$1.24\times\sqrt{I^2+J^2}$\nl
\nl
&    & {\it K. Zero Point Error in $V$-band\/} && $\pm 0.02$ mag\nl
&    & {\it L. Error in $(V-I_c)$\/} && $\pm 0.02$ mag\nl
&    & {\it M. Transformation to Gunn $r$\/}  && $\pm 0.02$ mag\nl
&    & {\it N. Error Due to Color Gradients\/} && $\pm 0.02$ mag\nl
&    & {\it O. Error in DIRBE Extinctions\/} && $\pm 0.02$ mag\nl
&S2.2& Systematic Photometric Error in Gunn $r$$^e$ && $\pm 0.04$ mag&
$\sqrt{(K^2+0.4^2L^2)+M^2+N^2+O^2}$\nl
\nl
&R2.1& Random Error in measurement of $\theta_e\langle I\rangle_e^{0.82}$,
$r_n$ && $\pm 0.02$ dex\nl
\nl
&R2.2$_L$& Random Error in measurement of zero points of FP and \dns
for Leo I&& $\pm 0.03$ dex & Due to finite number of galaxies in the
sample \nl
&R2.2$_V$& Random Error in measurement of zero points of FP and \dns
for Virgo&& $\pm 0.01$ dex\nl
&R2.2$_L$& Random Error in measurement of zero points of FP and \dns
for Fornax&& $\pm 0.03$ dex\nl
&R2.2& Internal random error in weighted mean zero point of FP and \dns &&
$\pm 0.01$ dex & Uncorrelated sum of R2.2$_L$, R2.2$_V$, and R2.2$_F$
\nl
\nl
&R2.3& Random error in zero point due to random errors in the Cepheid scale
&& $\pm 0.01$ dex & Uncorrelated sum of R1$_L$, R1$_V$, and R1$_F$
\nl
\nl
&S2& Total systematic error in nearby fundamental plane and \dns\ zero point
&& $\pm 0.03$ dex & Uncorrelated sum of S2.1 and S2.2
\nl
&R2& Total random error in nearby fundamental plane and \dns\ zero point
&& $\pm 0.02$ dex & Uncorrelated sum of R2.1, R2.2, and R2.3
\nl
\nl

\multicolumn{5}{l}{\bf 3. Errors in the Hubble Constant}\nl
&R3.1& Random error due to peculiar velocities of distant clusters, per cluster&& $\pm 1\%$
& Estimated error is $4\%/\sqrt{N}$, where $N=11$ clusters\nl
&R3.2& Random error due to uncertainties in FP and \dns\ Slopes&& $\pm 3\%$\nl

&S3.1& Systematic correction due to Cepheid Metallicity Correction&&
$-6\%\pm4\%$ & Using $\Delta \mu_{VI_c}=-0.24$ mag/dex [O/H]\nl

&S3.2& Systematic error due to cluster population incompleteness&&
$\pm 2\%$\nl

&S3.3& Systematic correction due to spatial coincidence of spirals and
ellipticals&& $-5\%\pm5\%$\nl

\nl
&R$_{H_0}$& Total Random Error in $H_0$ && $\pm 6\%$ \nl
&S$_{H_0}$& Total Systematic Error in $H_0$ && $\pm 11\%$ \nl
\enddata
\tablecomments{
$^a$Errors are in magnitudes for Sections 1., and as
indicated for Sections 2. and 3.
$^b$Equal to the scatter in the de-reddened PL relation for the LMC,
divided by $\sqrt{N}$.
$^c$Contributing uncertainties from the Holtzman \etal\ (1995)
zero points and the long-short uncertainty, combined in quadrature.
$^d$Values quoted are typical of the galaxies in the Key Project, but
individual cases vary. See the individual references for the appropriate
values for specific galaxies.
$^e$The coefficient on $L$ is simply the color term of the transformation
in \S \ref{transform}.
$^f$Uncertainty in DIRBE reddening taken from Schlegel \etal\ (1998).
}
\end{deluxetable}
\clearpage


\begin{deluxetable}{l c c c c c}
\tiny
\tablewidth{0pt}
\tablecaption{Mean zero points for the Fundamental Plane and
\dns\ Relations\label{tab:mzpts}}
\tablehead{
\colhead{Cluster} &
\colhead{$m-M$$^a$ (mag)} &
\colhead{$\langle \gamma\rangle $$^b$} &
\colhead{$\langle \delta\rangle $$^c$} &
\colhead{Cepheid Random (dex)} &
\colhead{Cepheid Systematic (dex)}
}
\startdata
Leo I  & $30.08\pm 0.11\pm 0.16$& $-0.108\pm 0.033$ & $-2.341\pm 0.033$ & $\pm 0.022$ & $\pm 0.032$\nl
Virgo  & $31.03\pm 0.04\pm 0.16$& $-0.182\pm 0.012$ & $-2.403\pm 0.012$ & $\pm 0.008$ & $\pm 0.032$\nl
Fornax & $31.60\pm 0.14\pm 0.16$& $-0.173\pm 0.033$ & $-2.388\pm 0.033$ & $\pm 0.028$ & $\pm 0.032$\nl
\nl
{\it weighted mean\/} & \nodata & $-0.173\pm 0.011$ & $-2.395 \pm 0.011$ & $\pm 0.007$ & $\pm 0.032$\nl
\enddata
\tablecomments{
$^a$Weighted mean distance moduli from Table \ref{tab:calibs} with
random and systematic uncertainties. The errors in the Cepheid
distance scale are detailed in Table \ref{tab:errors}\nl
$^b$$\gamma=\log r_e-1.24\log\sigma+0.82\log\langle I\rangle_e$.
Internal random errors given.\nl
$^c$$\delta=\log r_n-1.24\log\sigma$. Internal random errors given.\nl
}
\end{deluxetable}


\begin{deluxetable}{l r c c c r c r c}
\tiny
\tablewidth{0pt}
\tablecaption{The Distant Cluster Sample \label{tab:allh0}}
\tablehead{
\colhead{Cluster} &
\colhead{N} &
\colhead{Zero Point$^a$} &
\colhead{\rms} &
\colhead{$D$ (Mpc)} &
\colhead{$cz_{\rm CMB}$} &
\colhead{$H_0$$^b$} &
\colhead{$cz_{\rm flow}$} &
\colhead{$H_0$$^b$}
}
\startdata
\cutinhead{Fundamental Plane}
 Dorado Cluster &  9 & -4.336 & 0.103 & $  15\pm  2$&  1131 & $  78\pm  8$&  1064 & $  73\pm  8$\nl
        Hydra I & 20 & -4.886 & 0.141 & $  52\pm  5$&  4061 & $  79\pm  8$&  3881 & $  75\pm  7$\nl
     Abell S753 & 16 & -4.892 & 0.098 & $  52\pm  5$&  4351 & $  83\pm  7$&  3973 & $  76\pm  7$\nl
     GRM 15$^c$ &  7 & -4.871 & 0.088 & $  50\pm  5$&  4530 & $  91\pm  9$&  4848 & $  97\pm 10$\nl
     Abell 3574 &  7 & -4.908 & 0.095 & $  54\pm  6$&  4749 & $  87\pm  9$&  4617 & $  85\pm  9$\nl
      Abell 194 & 25 & -4.942 & 0.093 & $  59\pm  5$&  5100 & $  87\pm  7$&  5208 & $  89\pm  7$\nl
     Abell S639 & 12 & -4.970 & 0.087 & $  63\pm  6$&  6533 & $ 104\pm  9$&  6577 & $ 105\pm  9$\nl
   Coma Cluster & 81 & -5.129 & 0.085 & $  90\pm  6$&  7143 & $  79\pm  6$&  7392 & $  82\pm  6$\nl
     DC 2345-28 & 30 & -5.205 & 0.079 & $ 108\pm  8$&  8500 & $  79\pm  6$&  8708 & $  81\pm  6$\nl
      Abell 539 & 25 & -5.204 & 0.068 & $ 107\pm  8$&  8792 & $  82\pm  6$&  8648 & $  80\pm  6$\nl
     Abell 3381 & 14 & -5.309 & 0.101 & $ 137\pm 12$& 11536 & $  84\pm  8$& 11436 & $  84\pm  8$\nl
\cutinhead{\dns}
 Dorado Cluster &  9 & -6.600 & 0.126 & $  16\pm  2$&  1131 & $  71\pm  8$&  1064 & $  66\pm  8$\nl
        Hydra I & 20 & -7.130 & 0.128 & $  54\pm  5$&  4061 & $  75\pm  7$&  3881 & $  71\pm  7$\nl
     Abell S753 & 16 & -7.145 & 0.108 & $  56\pm  5$&  4351 & $  77\pm  7$&  3973 & $  71\pm  6$\nl
     GRM 15$^c$ &  7 & -7.092 & 0.086 & $  50\pm  5$&  4530 & $  91\pm  9$&  4848 & $  97\pm 10$\nl
     Abell 3574 &  7 & -7.142 & 0.107 & $  56\pm  6$&  4749 & $  85\pm 10$&  4617 & $  83\pm  9$\nl
      Abell 194 & 25 & -7.178 & 0.096 & $  61\pm  5$&  5100 & $  84\pm  7$&  5208 & $  86\pm  7$\nl
     Abell S639 & 12 & -7.224 & 0.085 & $  68\pm  6$&  6533 & $  97\pm  9$&  6577 & $  97\pm  9$\nl
   Coma Cluster & 81 & -7.373 & 0.084 & $  95\pm  7$&  7143 & $  75\pm  5$&  7392 & $  78\pm  5$\nl
     DC 2345-28 & 30 & -7.441 & 0.081 & $ 111\pm  8$&  8500 & $  76\pm  6$&  8708 & $  78\pm  6$\nl
      Abell 539 & 25 & -7.445 & 0.076 & $ 112\pm  8$&  8792 & $  78\pm  6$&  8648 & $  77\pm  6$\nl
     Abell 3381 & 14 & -7.540 & 0.085 & $ 140\pm 12$& 11536 & $  83\pm  7$& 11436 & $  82\pm  7$\nl
\enddata
\tablecomments{
$^a$These mean fundamental plane and \dns\ zero points are given by
$\gamma=\log \theta_e -1.24 \log \sigma +0.82 \log \langle I\rangle_e$ and
$\gamma=\log \theta_n -1.24 \log \sigma$, respectively, where
$\theta_e$ and $\theta_n$ are measured in radians.\nl
$^b$The uncertainties listed are the quadrature sum of the random errors
in the local, Cepheid-based zero points, the random errors in zero point
for each distant cluster, and the random error due to uncertainties in
the FP and \dns slopes.
$^c$Grm 15 is the designation by J\o{}rgensen \etal\ (1995) for southern group
number 15 of Maia, da Costa, \& Latham (1989).
}
\end{deluxetable}


\begin{thebibliography}{referencelist}

\bibitem[]{irtf}Aaronson, M., Mould, J., \& Huchra, J. 1979, \apj,
229, 1

\bibitem[]{bend}Bender, R., Saglia, R.P., \& Gerhard, O.E. 1994,
\mnras, 269, 785

\bibitem[Bower, Lucey, \& Ellis 1992]{bower}Bower, R.G., Lucey, J.R., \&
Ellis, R.S. 1992, \mnras, 254, 601

\bibitem[Burstein, Faber, \& Dressler 1990]{bfd}Burstein, D., Faber, S.M.,
\& Dressler, A. 1990, \apj, 354, 18

\bibitem[Caon, Capaccioli, \& D'Onofrio 1993]{caon}Caon, N., Capaccioli, M.,
\& D'Onofrio, M. 1993, \mnras, 265, 1013

\bibitem[Cardelli, Clayton, \& Mathis 1989]{ccm}Cardelli, J.~A., Clayton,
G.~C., \& Mathis, J.~S. 1989, \apj, 345, 245

\bibitem[Ciotti \& Lanzoni 1997]{ciotti2}Ciotti, L. \& Lanzoni, B., 1997,
A\&A, 321, 724

\bibitem[Colless \& Dunn 1996]{coma}Colless, M., \& Dunn, A.M. 1996,
\apj, 458, 435

\bibitem[davies1]{davies1}Davies, R., 1981, \mnras, 194, 879

\bibitem[davies2]{davies87}Davies, R., Burstein, D., Dressler, A.,
Faber, S.M., Lynden-Bell, D., Terlevich, R.H. \& Wegner, G. 1987,
\apjs, 64, 581

\bibitem[de Vaucouleurs 1961]{colg}de Vaucouleurs, G. 1961, \apjs, 5,
233

\bibitem[de Vaucouleurs \& Olson 1982]{dvolson}de Vaucouleurs, G. \& Olson,
D. 1982, \apj, 256, 346

\bibitem[Djorgovski \& Davis 1987]{dd87}Djorgovski S., \&  Davis M. 1987,
\apj, 313, 59

\bibitem[Djorgovski, de Carvalho, \& Han 1988]{djorgeg}Djorgovski, S.,
de Carvalho, R. \& Han, M.-S. 1988, in ``The Extragalactic Distance
Scale'' ed. S. van den Bergh \& C.J. Pritchet (ASP Conf. Ser., 4), 329

\bibitem[]{ital}D'Onofrio, M., Zaggia, S.R., Longo, G., Caon, N.,
Capaccioli, M. 1995, A\&A, 296, 319

\bibitem[]{ital2}D'Onofrio, M., Capaccioli, M., Zaggia, S.R., \& Caon,
N. 1997, \mnras, 289, 847

\bibitem[Dressler 1984]{dress84}Dressler, A. 1984, \apj, 281, 512

\bibitem[Dressler \etal\ 1987]{dress7s}Dressler, A., Lynden-Bell, D.,
Burstein, D., Davies, R.L., Faber, S.M, Terlevich, R.J. \& Wegner, G. 1987,
\apj, 313, 42

\bibitem[Faber \& Jackson 1976]{fj76}Faber, S.M. \& Jackson, R.E. 1976,
\apj, 204, 668

\bibitem[Faber \etal\ 1987]{faber87}Faber S. M., Dressler A., Davies R. L.,
Burstein D., Lynden-Bell D., Terlevich R., \&  Wegner G. 1987, Faber S. M.,
ed., Nearly Normal Galaxies. Springer, New York, p. 175

\bibitem[]{f89}Faber, S.M., Wegner, G., Burstein, D., Davies, R.L.,
Dressler, A., Lynden-Bell, D., \& Terlevich, R.H. 1989, \apjs, 69, 763

\bibitem[]{cpib}Federspiel, M., Sandage, A. \& Tammann, G.A. 1994,
\apj, 430, 29

\bibitem[]{f96}Ferrarese, L., \etal\ 1996, \apj, 464, 568

\bibitem[]{f99}Ferrarese, L., \etal\ 1999, \apj, accepted for
publication

\bibitem[]{fish}Fisher, D. 1997, AJ, 113, 950

\bibitem[]{freed}Freedman, W. L., \etal\ 1999, \apj, in preparation

\bibitem[Frei \& Gunn 1994]{frei}Frei, Z., \& Gunn, J.E. 1994, \aj, 108,
1476

\bibitem[]{f93}Fukugita, M.,Hogan, C.J., \& Peebles, P.J.E. 1993,
Nature, 366, 309

\bibitem[]{rgfp}Gibbons, R.A, Fruchter, A.S., \& Bothun, G.D. 1998,
astro-ph/9903380

\bibitem[]{g99}Gibson, G. 1999, \apj, accepted for publication

\bibitem[Graham \& Colless 1997]{graham}Graham, A., \& Colless, M. 1997,
\mnras, 287, 221

\bibitem[]{gonz}Gonzalez, A.H., \& Faber, S.M. 1997, \apj, 485, 80

\bibitem[]{john}Graham, J., \etal\ 1997, \apj, 477, 535

\bibitem[]{john2}Graham, J., \etal\ 1999, \apj, in press

\bibitem[Gregg 1992]{gregg2}Gregg, M.D. 1992, \apj, 384, 43

\bibitem[]{gclf}Hanes, D.A., \& Whittaker, D.G. 1987, \aj, 94, 906

\bibitem[]{ht}Hjorth, J. \& Tanvir, N.R. 1997, \apj, 482, 68

\bibitem[]{holtz}Holtzman, J.A., \etal\ 1995, \pasp, 107, 1065

\bibitem[]{hudson}Hudson, M.J., Lucey, J.R., Smith, R.J., \& Steel J.
1997, \mnras, 291, 488

\bibitem[]{j92}Jacoby, G.H. 1989, \apj, 339, 39

\bibitem[]{jetal92}Jacoby, G.H., \etal\ 1992, \pasp, 104, 599

\bibitem[J\o{}rgensen \etal\ 1992]{jfk92}J\o{}rgensen I., Franx M., \&
Kj\ae{}rgaard P. 1992, A\&AS, 95, 489

\bibitem[J\o{}rgensen 1994]{j94}J\o{}rgensen I. 1994, \pasp, 106, 967

\bibitem[J\o{}rgensen \etal\ 1993]{jfk93}J\o{}rgensen I., Franx M., \&
Kj\ae{}rgaard P. 1993, \apj, 411, 34

\bibitem[J\o{}rgensen \etal\ 1995a]{jfk95a}J\o{}rgensen I., Franx M., \&
Kj\ae{}rgaard P. 1995a, \mnras, 273, 1097

\bibitem[J\o{}rgensen \etal\ 1995b]{jfk95b}J\o{}rgensen I., Franx M., \&
Kj\ae{}rgaard P. 1995b, \mnras, 276, 1341

\bibitem[J\o{}rgensen \etal\ 1996]{jfk96}J\o{}rgensen I., Franx M., \&
Kj\ae{}rgaard P. 1996, \mnras, 280, 167 [JFK96]

\bibitem[J\o{}rgensen 1998]{j99}J\o{}rgensen I. 1999, private
communication

\bibitem[]{thesis}Kelson, D.D., 1998, Ph.D. thesis, Univ. of
California, Santa Cruz

\bibitem[]{k99}Kelson, D.D., 1999, \apj, in preparation

\bibitem[]{k99a}Kelson, D.D., \etal\ 1999a, \apj, in press

\bibitem[]{k99b}Kelson, D.D., \etal\ 1999b, \apj, submitted

\bibitem[]{k99c}Kelson, D.D., \etal\ 1999c, \apj, submitted

\bibitem[]{metal}Kennicutt, R.C., \etal\ 1998, \apj, 498, 181

\bibitem[]{landolt}Landolt, A., 1992, \aj, 104, 372

\bibitem[Lucey, Bower, \& Ellis 1991]{lucey}Lucey, J.R., Bower, R.G., \&
Ellis, R.S. 1991, \mnras, 249, 755

\bibitem[Lynden-Bell \etal\ 1987]{lynbel7s}Lynden-Bell, D., Faber,
S.M, Burstein, D., Davies, R.L., Dressler, A., Terlevich, R.J. \&
Wegner, G. 1987, \apj, 326, 19

\bibitem[]{macri}Macri, L.M., \etal\ 1999, \apj, in press

\bibitem[]{maia}Maia, M.A.G., da Costa, L.N., Latham, D.W. 1989, \apjs,
69, 809

\bibitem[]{mf}Madore, B.F. \& Freedman, W.L. 1999, \apj\, in preparation

\bibitem[]{m99}Mould, J.R. 1999a, \apj, submitted

\bibitem[]{m99b}Mould, J.R. 1999b, \apj, submitted

\bibitem[Pahre 1998]{pahrethesis}Pahre, M.A., 1998, Ph.D. thesis,
California Inst. of Technology

\bibitem[]{p99}Prosser, C., \etal\ 1999, \apj, submitted

\bibitem[Renzini \& Ciotti 1993]{renzini}Renzini, A., \& Ciotti, L. 1993,
\apj, 416, 49

\bibitem[]{sn96}Riess, A.G., Press, W.H., \& Kirshner, R.P. 1996,
\apj, 473, 88

\bibitem[Saglia \etal\ 1997]{saglia}Saglia, R.P., \etal\ 1997, \apjs, 109,
79

\bibitem[Sandage \& Visvanathan 1978]{sandagecm}Sandage, A.R., \&
Visvanathan, N. 1978, \apj, 225, 742

\bibitem[]{s99}Sakai, S. 1999, \apj, accepted for publication

\bibitem[Schlegel \etal\ 1998]{Dust98}Schlegel, D.J., Finkbeiner,
D.P., \& Davis, M. 1998,
\apj, 500, 525

\bibitem[Sersic 1968]{sersic}Sersic, J.L. 1968, {\it Atlas de Galaxia
Australes\/}, (Cordoba: Observatorio Astronomico)

\bibitem[]{nancy}Silbermann, N., \etal\ 1999, \apj, 515, 1

\bibitem[Suntzeff \etal\ 1999]{suntz}Suntzeff, N. B., \etal\ 1999, AJ,
in press

\bibitem[Terlevich \etal\ 1981]{terlevich}Terlevich, R.J., Davies, R.L.,
Faber, S.M., \& Burstein, D. 1981, MNRAS, 196, 381

\bibitem[]{tg76}Thuan T.X., \& Gunn, J., 1976, \pasp, 88, 543

\bibitem[]{t81}Tonry, J., \& Davis, M. 1981, \apj, 246, 680

\bibitem[]{t83}Tonry, J. 1983, \apj, 266, 58

\bibitem[]{t88}Tonry, J., \& Schneider, D.P. 1988, \aj, 96, 807

\bibitem[]{t97}Tonry, J., Blakeslee, J.P., Ajhar, E.A., \& Dressler,
A. 1997, \apj, 475, 399

\bibitem[]{t99}Tonry, J., Blakeslee, J.P., Dressler, A.,
\& Ajhar, E.A., 1999, in preparation

\bibitem[Trager 1997]{trager}Trager, S.C. 1997, Ph.D. thesis, Univ. Calif.,
Santa Cruz

\bibitem[Tully \& Fisher 1977]{tf77}Tully, R.B. \& Fisher J.R. 1977, A\&A,
54, 661

\bibitem[]{v98}van Dokkum, P.G., Franx, M., Kelson, D.D., \&
Illingworth, G.D. 1998, \apjl, 504, 17

\bibitem[]{w98}Wegner, G., Colless, M., Baggley, G., Davies, R.L,
Bertschinger, E., Burstein, D., McMahan, R.K., Jr., \& Saglia, R.P.
1996, \apjs, 106, 1

\bibitem[Worthey, Trager, \& Faber 1996]{wtf}Worthey, G., Trager, S. C., \&
Faber, S. M. 1996, in ``Fresh Views on Elliptical Galaxies,'' eds. A.
Buzzoni, A. Renzini, \& A. Serrano, (ASP Conf. Ser., 86) p. 203 

\end{thebibliography}
\end{document}